\documentclass[10pt]{article}
\usepackage[a4paper, total={7in, 9in}]{geometry}
\usepackage[utf8]{inputenc}
\usepackage{graphicx}
\usepackage[dvipsnames]{xcolor}
\usepackage{amsmath,amssymb}
\usepackage{soul}
\usepackage{multirow}
\usepackage{rotating}
\usepackage{enumitem}
\usepackage{authblk}
\usepackage{hyperref}
\usepackage{xspace}
\usepackage{booktabs}% per \tabitem
\usepackage[section]{placeins} % per \FloatBarrier

\newcommand{\Rt}{\ensuremath{R_t}\xspace}
\newcommand{\Rtstar}{\ensuremath{R_t^*}\xspace}
\newcommand{\Rtfit}{\ensuremath{\hat{R}_t}\xspace}
\newcommand{\tdelay}{\ensuremath{t_{D}}\xspace}

\begin{document}

\title{Study on the effects of the restrictive measures for containment of the COVID-19 pandemic on the reproduction number \Rt in Italian regions}

\author[1,2]{Gianluca Bonifazi}
\author[3,4,5]{Luca Lista}
\author[6]{Dario Menasce}
\author[7,*]{Mauro Mezzetto}
\author[6]{Daniele Pedrini}
\author[2]{Roberto Spighi}
\author[8,2]{Antonio Zoccoli}
\affil[1]{\raggedright\normalsize Universit\`a Politecnica delle Marche}
\affil[2]{\raggedright\normalsize INFN Sezione di Bologna}
\affil[3]{\raggedright\normalsize Universit\`a degli Studi di Napoli Federico II}
\affil[4]{\raggedright\normalsize INFN Sezione di Napoli}
\affil[5]{\raggedright\normalsize Scuola Superiore Meridionale}
\affil[6]{\raggedright\normalsize INFN Sezione di Milano Bicocca}
\affil[7]{\raggedright\normalsize INFN Sezione di Padova}
\affil[8]{\raggedright\normalsize Alma Mater Studiorum Universit\`a di Bologna}
\affil[*]{\raggedright\normalsize \textbf{Corresponding author}, e-mail: {\tt mauro.mezzetto@pd.infn.it}}
\date{}

\maketitle

\begin{abstract}
Since November 6$^{\mathrm{th}}$, 2020, Italian regions have been classified according to four levels, corresponding to specific risk scenarios, for which specific restrictive measures have been foreseen. By analyzing the time evolution of the reproduction number \Rt, we estimate how much different restrictive measures affect \Rt, and we quantify the combined effect of the diffusion of virus variants and the beginning of the vaccination campaign upon the \Rt trend. We also compute the time delay between implementation of restrictive measures and the resulting effects. Three different models to describe the effects of restrictive measures are discussed and the results are cross-checked with two different algorithms for the computation of \Rt.
\end{abstract}

\section{Introduction}
The main purpose of this paper is to quantify the effects on the reproduction number \Rt of the Non-Pharmaceutical Interventions (NPI) put in place in the Italian regions since November 6$^{\mathrm{th}}$, 2020. 

\begin{figure}[htbp]
    \centering
    \includegraphics[width=\textwidth]{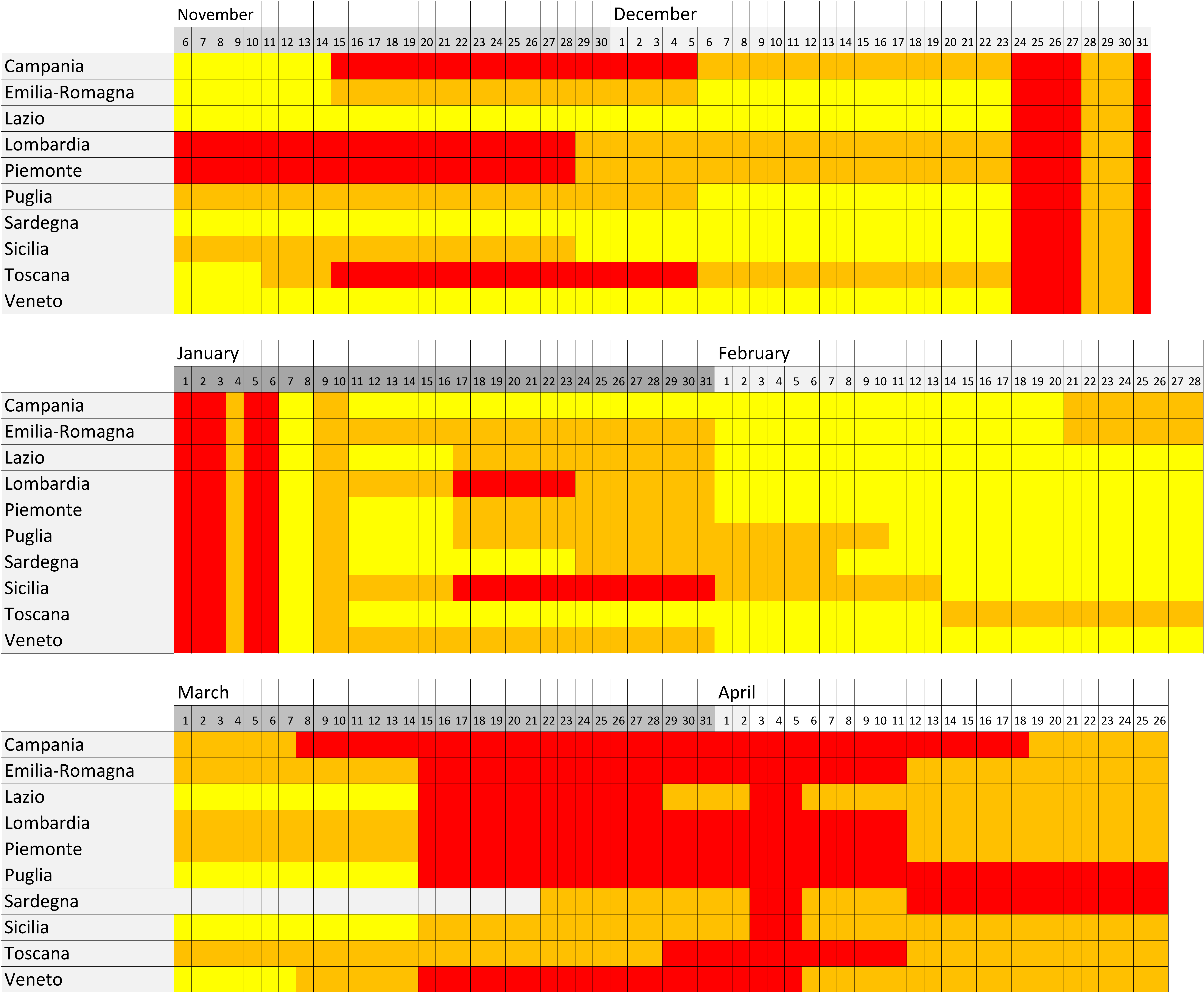}
    \caption{\footnotesize NPI periods in the Italian regions, characterized by the colored markers white, yellow, orange and red, in the November 6$^{\mathrm{th}}$, 2020 - April 26$^{\mathrm{th}}$, 2021 period of time.  Besides the nine regions previously mentioned, it is also reported Sardegna, the only Italian region to reach the white restriction in the time period considered.} 
    \label{fig:colori}
\end{figure}

Italian regions and the autonomous provinces of Trento and Bolzano have been classified into the four aforementioned colored levels: red, orange, yellow and white in decreasing levels of restrictions, corresponding to four risk scenarios, for which specific restrictive measures were foreseen, as reported in the Appendix. %~\ref{app:risks}. 
The scenarios are described in \cite{scenarios} and risk indicators are defined in the Ministerial Decree of April 30$^{\mathrm{th}}$, 2020 \cite{risk-indicators}. 
The color assignments to the Italian regions along the period November 6$^{\mathrm{th}}$, 2020 -- April 26$^{\mathrm{th}}$, 2021 are displayed in Figure~\ref{fig:colori}. We note that the change of NPI  in this period was quite frequent. For instance, Lombardia, the most populated Italian region with about 10.0 million inhabitants, changed level 15 times in about six months.

Three possible models to evaluate the effects of the  NPI  on \Rt will be compared, and we will address two aspects that affect its computation:
\begin{itemize}
\item how  can \Rt be properly synchronized with the daily number of positive swabs, from which it is computed;
\item the value of the time delay between the implementation of a new level of NPI and its effects on \Rt.
\end{itemize}

We will compare the effects of the different NPI  in the nine most populated Italian regions:  Lombardia, Lazio, Campania, Sicilia, Veneto, Emilia-Romagna, Piemonte, Puglia, Toscana (in decreasing level of population). All together, these regions contain  80\% of the total Italian population. We excluded regions with less than 2 million inhabitants, in order to ensure a sufficiently large number of cases and allow a precise \Rt computation.

In early January 2021 the vaccination campaign started in Italy, and in the same period novel  virus variants spread across the country.
These two important factors potentially affect \Rt in opposite ways and we  will compute their combined effect.

% ---------------------------------------------------------------------
%-----------------------------------------------------------------------------------------------
\section{$\mathrm{R_{t}}$ and its synchronization with the new COVID-19 cases}
\label{sec:Rt}
\subsection{Computation of $\mathrm{R_{t}}$}
\label{sec:comp-Rt}
We compute \Rt from the number of new COVID-19 daily cases reported by the Italian Dipartimento della Protezione Civile \cite{dpc}. In reference \cite{ourpaper}  (CovidStat from here on) we developed a simple, yet precise, algorithm to compute \Rt.
We publish daily real time estimates of \Rt, together with other indicators of the development of the Italian outbreak in \cite{our web site}. 

The official \Rt used by the Italian Government is computed from the date of onset of symptoms, not from the date of the positive swab  as has been done in this paper. We verified in \cite{sintomatici} that the two methods converge to essentially identical values of \Rt (see in particular Fig.~6 of reference \cite{sintomatici}), once the two estimations of \Rt are synchronized.
Furthermore, since the \Rt used by the Italian Government is computed with the Cori et al. algorithm \cite{Cori} (Cori from here on), we verified in \cite{ourpaper} and \cite{sintomatici} that our Covidstat algorithm computes \Rt in full agreement with it. In any case most of the results of this paper will be cross-checked with the Cori  algorithm, by using the {\tt EpiEstim} package~\cite{Epiestim} and assuming the same modeling of the generation time distribution as measured in~\cite{Cereda}.

We prefer to keep our Covidstat algorithm as the default for the following computations because it's much faster to compute (as discussed in \cite{ourpaper}) and because we have full control of its computation details in all the necessary steps, in particular when we proceed to synchronize \Rt with the data.

The \Rt computation  with the Covidstat algorithm requires  a smoothing of the original data as a preliminary step and, subsequently, an exponential fit to the   smoothed data, determining in this way the time-dependent growth rate $\lambda(t)$ of the number of new daily cases. 
The smoothing is performed applying the Savitzky--Golay algorithm~\cite{savitzky} in a time window of 21 days and with a second-order polynomial\footnote{The effect is very similar to a moving average, with the further advantage of being able to also perform the smoothing  on both edges of the considered time period.}.
The fit to an exponential curve is performed within  moving windows of 14 days.

\Rt is determined from  $\lambda=\lambda(t)$ using the  equation \ref{eq:rt_lista_3}:
\begin{equation}
    \Rt = (1+\lambda\theta)^\kappa\,\,,
    \label{eq:rt_lista_3}
\end{equation}

where $\theta$ and $\kappa$ are, respectively, the scale and  shape parameters of the gamma distribution that models the distribution of the generation time, the time interval between infector-infected pair~\cite{ourpaper}.  Estimates of the parameters $\theta$ and $\kappa$ of the generation time distribution are taken from~\cite{Cereda}.   

\subsection{Synchronization of $\mathrm{R_{t}}$}
\label{sec:synch}
To study the time-dependent effects of NPI  upon the trends of \Rt, it is important to assess the correct synchronization of the latter with the daily number of new  cases.
In real-time applications, it's customary to assign both the smoothing and the resulting $\lambda$ of the exponential fit to the last day of the time sequence. In this way, the value of \Rt can be assigned to the latest day.
A more reasonable option for synchronization would be to assign the smoothing and the parameter of the exponential fit to the central day of the sequence.

The way the CovidStat algorithm is defined,  provides an absolute mean to determine the optimal synchronization of \Rt: when the new cases are at a maximum or at a minimum, the growth rate $\lambda$ becomes zero and, according to Eq.\ref{eq:rt_lista_3}, \Rt must therefore be equal to one. This feature has the additional advantage of being independent from the assumed distribution of the generation time.

In order to verify the synchronization of \Rt with the number of new daily cases, we  determine the positions of the three prominent peaks in the period of time considered in this paper. To model them, we perform a fit to the sum of three Gompertz derivatives.  Gompertz curves are widely and commonly used to describe the development of pandemics, as discussed for instance in \cite{Gomp}.  Each derivative is defined  as:
\begin{equation}
    g(t;a,k_G,T_p) = a \cdot k_{\tiny G} e^{-e^{-k_{\tiny G}(t-T_p)}-k_{\tiny{G}}(t-T_p)}\,\,\,.
    \label{eq:gomp}
\end{equation}
In this parametrization, $a$ is the integral of the curve, $T_p$ it's the time when the peak is reached and $k_{\tiny G}$ is a growth-rate coefficient, modelling the shape, see also reference \cite{Gomp}.

\begin{figure}[htb]
\centering{
    \includegraphics[width=0.90\textwidth]{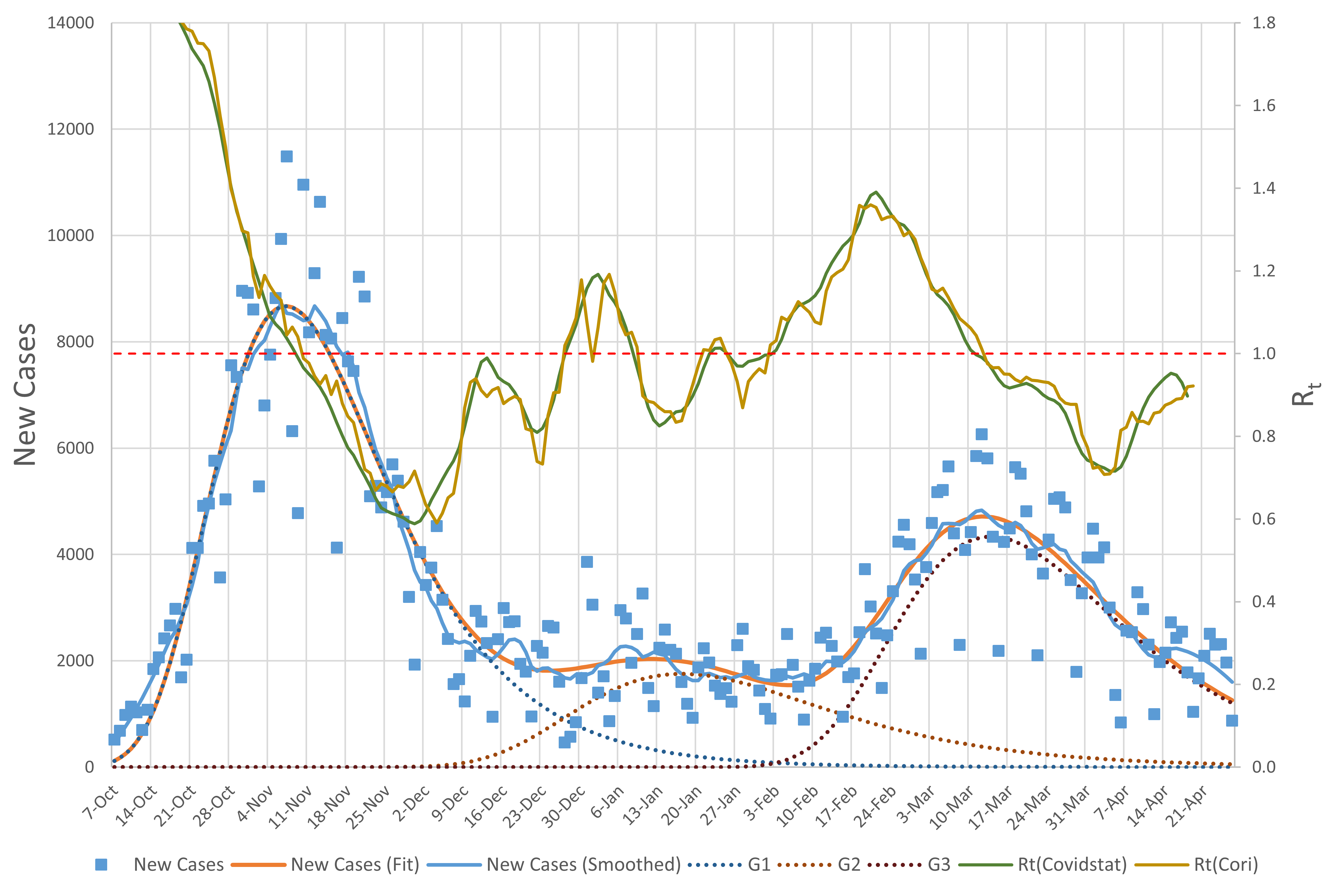}
    \caption{\footnotesize Distribution of the number of new daily cases in Lombardia (squares) together with their smoothing (blue line) and the fit to the sum of three Gompertz functions (orange line). Dotted curves indicate the three single Gompertz  components called G1, G2, G3. The \Rt values, as computed with the Covidstat algorithm (green curve) and with the Cori  algorithm (brown curve)  are superimposed with a different scale indicated  on the right of the plot. The dashed horizontal orange line is drawn to mark the critical value of \Rt=1.}}
    \label{fig:Lombardy}
\end{figure}

To illustrate the procedure, Figure~\ref{fig:Lombardy} shows the distribution of the number of new daily cases in Lombardia. A smoothed distribution and the fit of the sum of Gompertz functions are superimposed on the data. We note the large dispersion of the number of new daily cases to be well beyond their expected Poisson statistical fluctuation. This dispersion makes  the smoothing of the data necessary and sometimes produces false small secondary peaks that in turn are reflected in the evolution of \Rt.

Table~\ref{tab:peaks} displays  the day in which the  fitted curve reached the first peak 
in the nine Italian regions along with the dates when  \Rt crossed the critical value of 1.
The curves in Figure~\ref{fig:Lombardy} and the data in Table~\ref{tab:peaks}  show a good agreement between the dates in which a maximum of the new cases distribution is reached and the dates when $\Rt=1$. However, we had to assign the \Rt values computed with Cori to the initial day of the 
7-days time window under which it is computed, in contrast with the indications of~\cite{Cori}.

By adopting this synchronization (for the Covidstat algorithm the smoothing and the exponential fits are referred to the central day of the considered time window, while for Cori \Rt it is referred to the first day of the time window),  the Covidstat \Rt can only be computed up to 8 days before the last day of available data, while Cori can be computed up to 7 days before the last day.

\begin{table}[bht]
    \renewcommand{\arraystretch}{1.1}
    \centering
 \begin{tabular}{l|c|cc|rr}
 Region & Peak of new cases & {$R_t=1$} & {$R_t=1$} &\multicolumn{1}{c}{$\Delta t$} & \multicolumn{1}{c}{$\Delta t$}  \\
          & (days) & (Covidstat) & (Cori) & \multicolumn{1}{c}{(Covidstat)} & \multicolumn{1}{c}{(Cori)}\\
 \hline
  Piemonte       & $259.9 \pm 0.1$ & $262.5 \pm 0.1$ & $264.2 \pm 0.1$ & $ 2.6  \pm 0.1 $ & $ 4.3 \pm 0.1 $\\
  Lombardia      & $257.9 \pm 0.1$ & $259.6 \pm 0.1$ & $261.0 \pm 0.2$ & $ 1.7  \pm 0.1 $ & $ 3.1 \pm 0.2 $\\
  Veneto         & $266.7 \pm 0.1$ & $265.2 \pm 0.3$ & $268.8 \pm 0.6$ & $ -1.5 \pm 0.3 $ & $ 2.1 \pm 0.6$\\
  Emilia-Romagna & $266.5 \pm 0.1$ & $267.7 \pm 0.1$ & $270.0 \pm 0.3$ & $ 1.2  \pm 0.1 $ & $ 3.5 \pm 0.3$\\
  Toscana        & $258.1 \pm 0.2$ & $261.2 \pm 0.2$ & $264.1 \pm 0.1$ & $ 3.1  \pm 0.3 $ & $ 6.0 \pm 0.2$\\
  Lazio          & $262.4 \pm 0.3$ & $265.6 \pm 0.2$ & $264.9 \pm 0.1$ & $ 3.2  \pm 0.4 $ & $ 2.5 \pm 0.3$\\
  Campania       & $257.5 \pm 0.2$ & $259.1 \pm 0.1$ & $258.7 \pm 0.8$ & $ 1.6  \pm 0.2 $ & $ 1.2 \pm 0.8$\\
  Puglia         & $277.5 \pm 0.2$ & $281.4 \pm 0.1$ & $282.4 \pm 0.2$ & $ 3.9  \pm 0.2 $ & $ 4.9 \pm 0.3$\\
  Sicilia        & $268.4 \pm 0.7$ & $268.0 \pm 0.1$ & $268.4 \pm 0.2$ & $ -0.4 \pm 0.7 $ & $ 0.0 \pm 0.7$\\
      \end{tabular}
    \caption{\footnotesize Fit values of the peak position of the  second wave in nine Italian regions and date when \Rt curve crosses the $\Rt=1$ value, according to the Covidstat and the Cori algorithms, obtained with a straight line fit to the five \Rt values around \Rt=1. The difference of times $\Delta t$ between the peak of the new cases and the day when $\Rt=1$ are also reported.
    Days are counted since February 24$^{\mathrm{th}}$, 2020, day 260 corresponds to November 9$^{\mathrm{th}}$, 2020.}
    \label{tab:peaks}
\end{table}

Data in Table~\ref{tab:peaks} indicate a residual mismatch  of a couple of days on average between the day in which a maximum is reached and \Rt crosses the value of 1. We decided not to adjust the synchronization for this amount of time, because it could be partially due to a bias induced by the fit to the Gompertz functions;  in any case the procedure detailed in section~\ref{sec:td} guarantees an overall correct synchronization.

% ---------------------------------------------
\subsection{\Rt data}\label{sec:Rtdata}
\begin{figure}[htb]
\centering
    \includegraphics[width=0.32\textwidth]{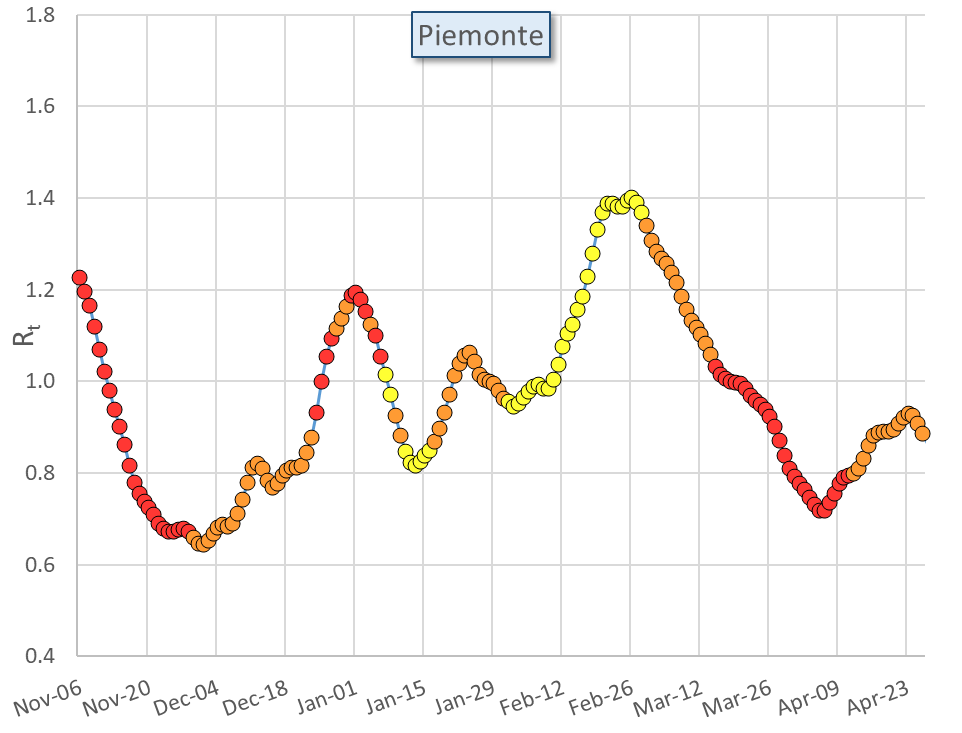}
    \includegraphics[width=0.32\textwidth]{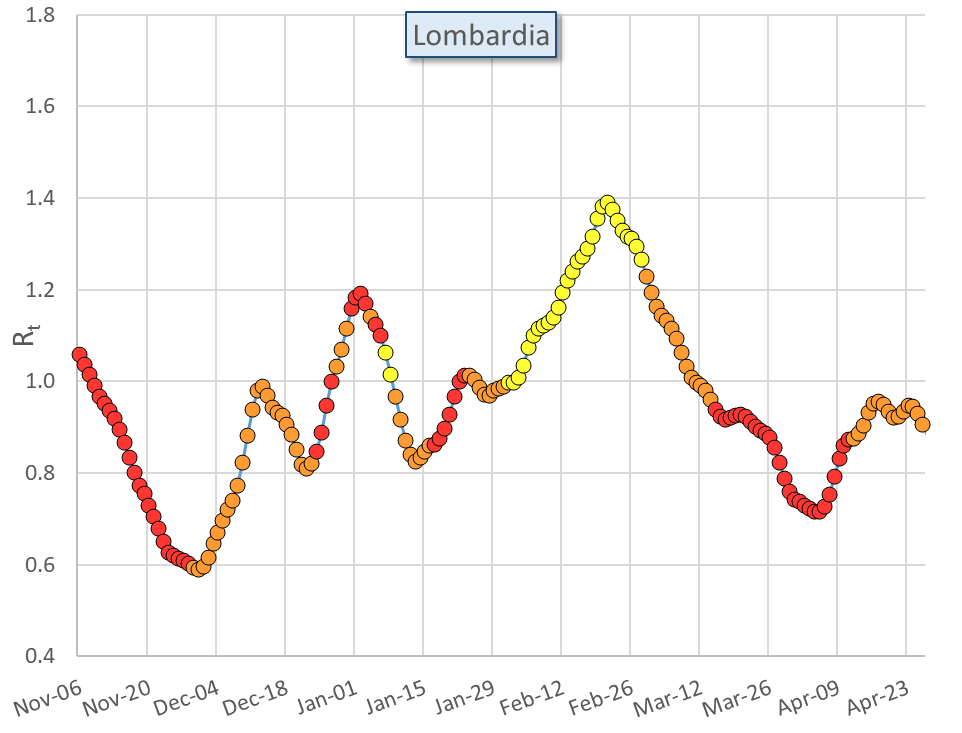}
    \includegraphics[width=0.32\textwidth]{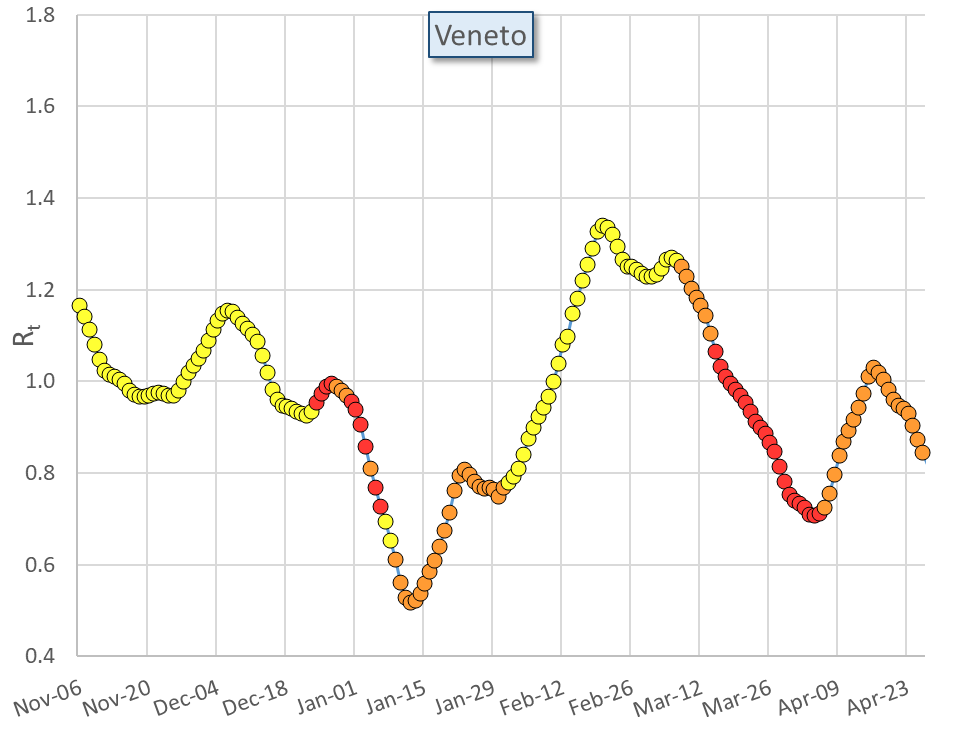}
    \includegraphics[width=0.32\textwidth]{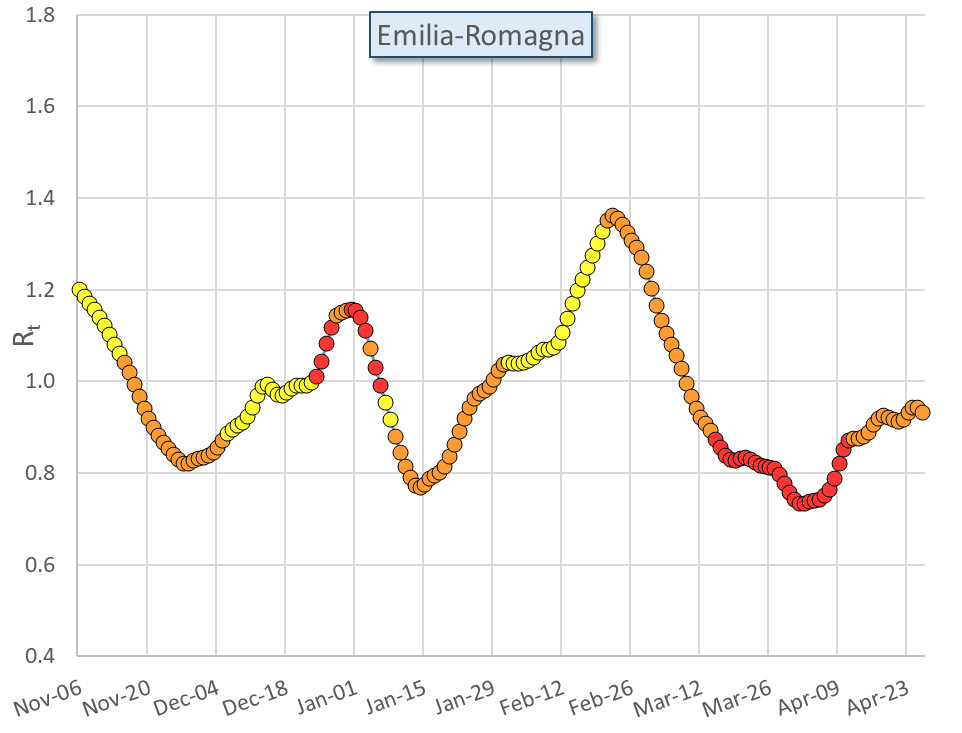}
    \includegraphics[width=0.32\textwidth]{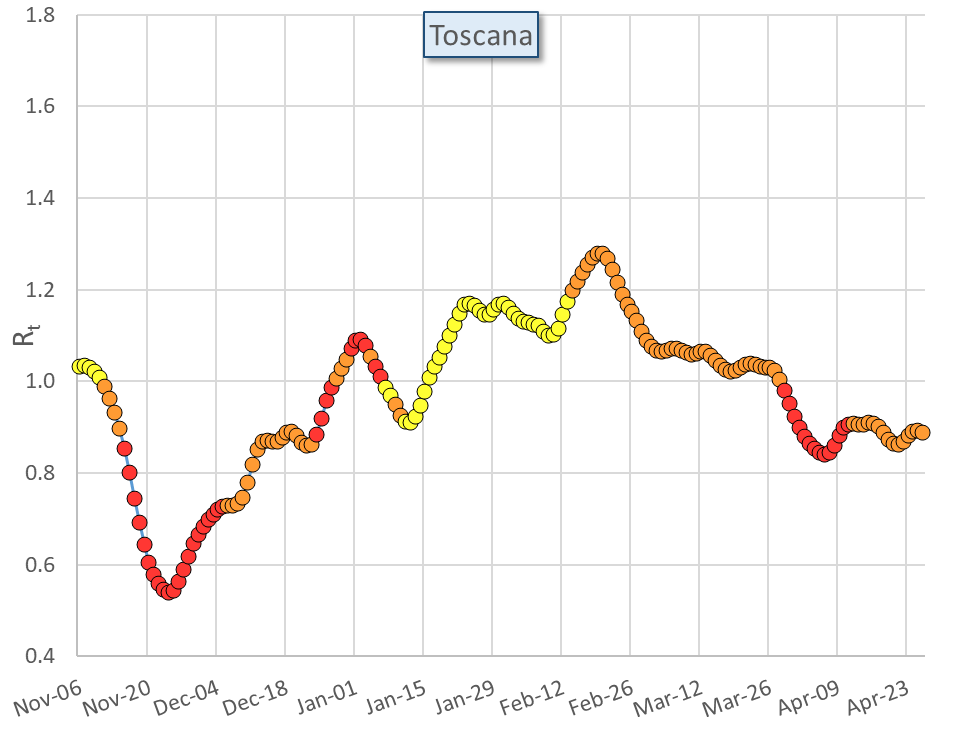}
    \includegraphics[width=0.32\textwidth]{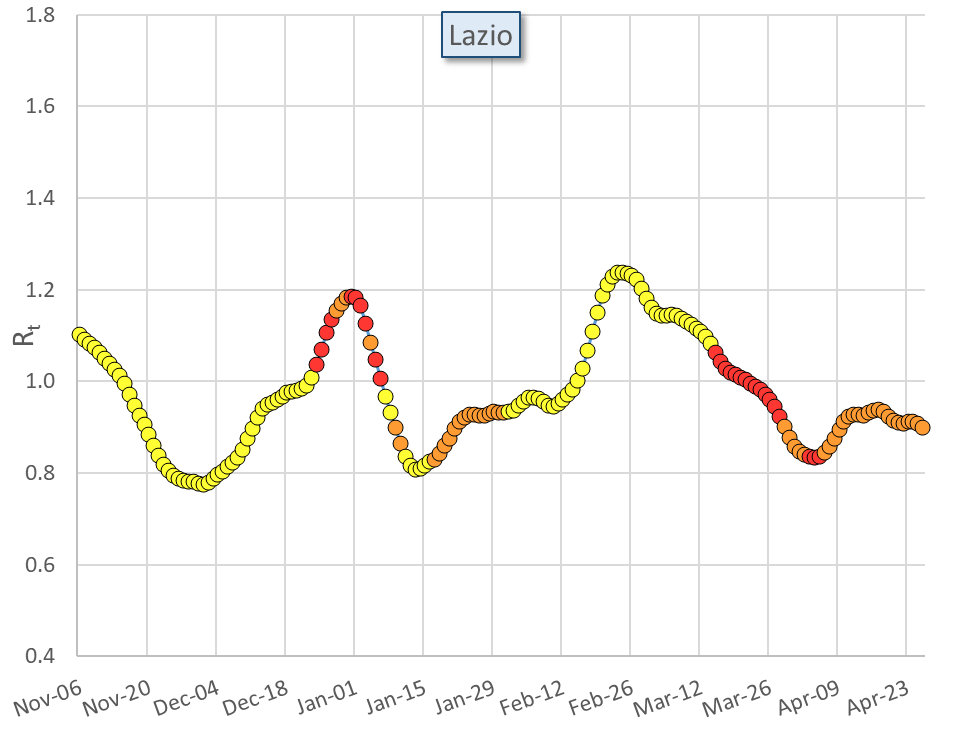}
    \includegraphics[width=0.32\textwidth]{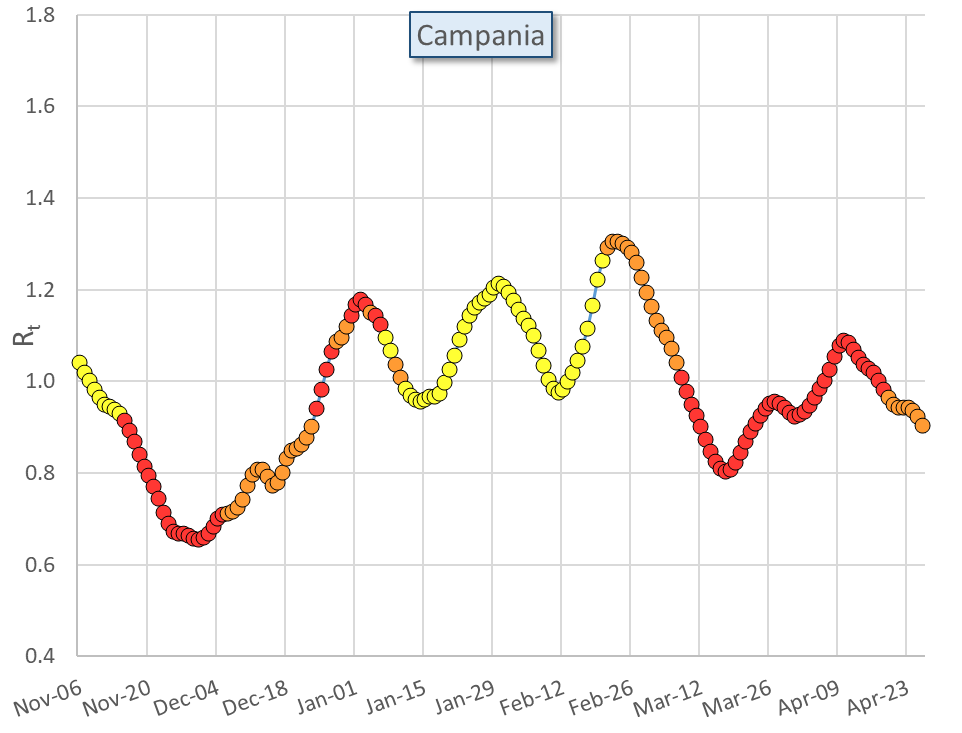}
    \includegraphics[width=0.32\textwidth]{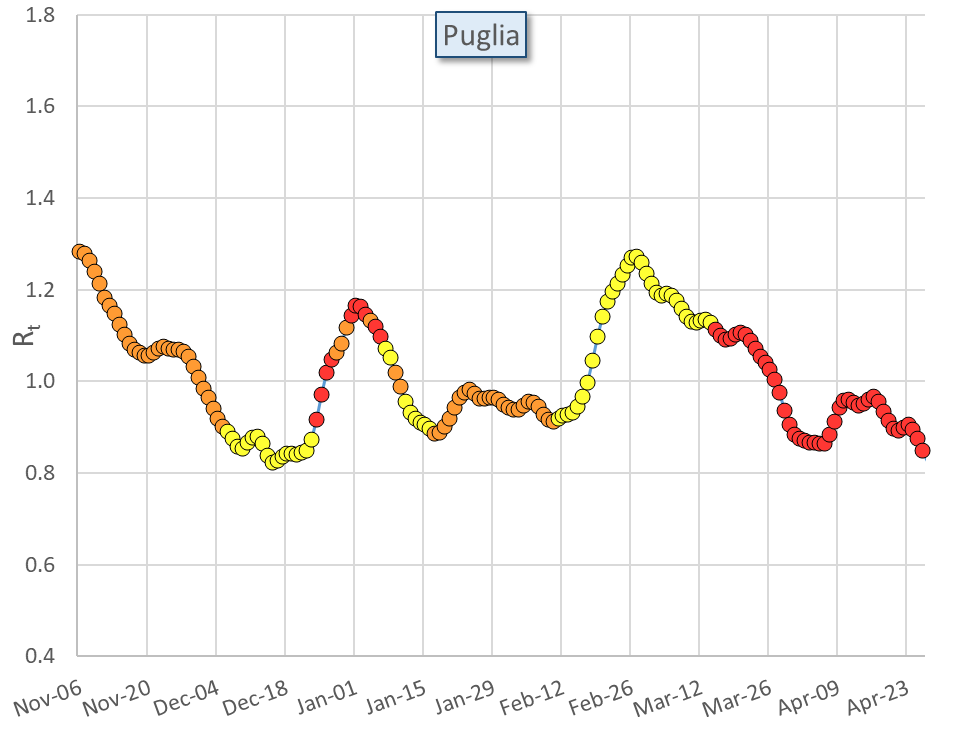}
    \includegraphics[width=0.32\textwidth]{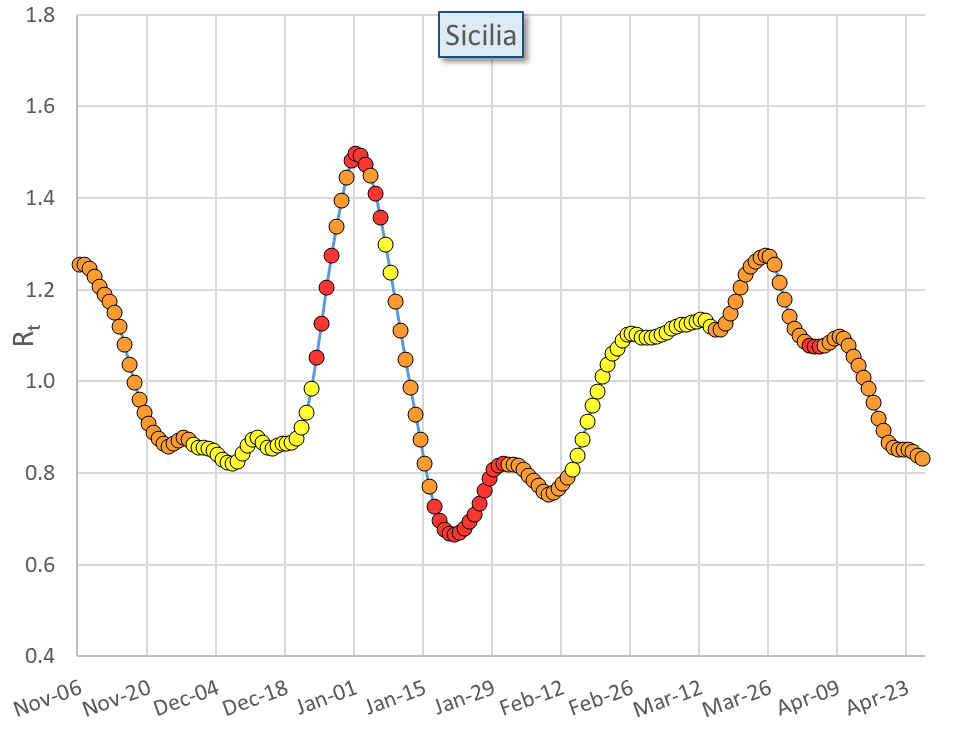}
    \caption{\footnotesize \Rt values as a function of time from November 6$^\mathrm{th}$, 2020 to April 26$^\mathrm{th}$, 2021 in the nine considered regions. The colors of the dots correspond to the three levels of the NPI (yellow, orange, red) locally enforced in different periods of time.}
    \label{fig:rt-ori}
\end{figure}

Once \Rt is synchronized with the number of new daily cases, we can proceed to plot the \Rt trends for the nine Italian regions considered.
Data are displayed in Figure~\ref{fig:rt-ori}, the colors of the dots correspond to the three NPI levels: yellow, orange and red.
None of the considered regions has ever been classified as white, so we can't compute the effects of the white NPI level.
We observe that all regions have the same descending trend at the beginning of the analyzed time range, because they were all subject, at the time, to the same  common NPI. Subsequently, the trends  began to differ from region to region, because of the uneven NPI  levels and their enforcement date. 

We have to consider several factors that can influence the \Rt behaviour beyond what is induced by the three NPI levels. Noticeably:
\begin{enumerate}[label=\alph*)]
\item the contagion tracking and screening procedures differ from region to region;  
\item individual regions  may have adopted specific additional restrictive provisions to those foreseen nation wide; in some occasions, limited parts of the regions, as single municipalities and provinces, where subject to stricter NPI;
\item  some degree of freedom about the  organization of schools of any degree along with Universities had been deferred to local level, more information about these points is reported in Appendix. Schools closures and partial openings, therefore, did not  follow the NPI levels in the same homogeneous way. Protocols for tracking students and schools staff and the actions to be taken in case of positive cases during the school openings were left to decisions taken at regional level.
\item mobility inside each region could significantly differ because of different work activities and different network of mobility connections;
\item the vaccination campaign started in January 2021 with a potential impact on \Rt development; 
\item virus variants appeared during the same period of time and became dominant towards its end  with a potential impact on \Rt development.
\end{enumerate}

For all these reasons, we believe that the effects induced by the NPI alone are not sufficient by themselves to fully account for the trends of \Rt. On the other hand, the purpose of this work is not to introduce a model capable of describing the development of the pandemic in all its aspects; we are rather interested in extracting the effects of the individual NPI.

% --------------------------------------------------------------------------------
\section{\Rt models}
 We  introduce three different models  to describe the effects of the NPI on the time evolution of \Rt.
They compute a predicted value for \Rt, \Rtfit, for all the  time periods in which the NPI  remained unchanged. At the beginning of each new time period we set $\Rtfit=\Rt$. The time periods do not begin on the calendar day in which the NPI  changed, $t$, but at the later day $t+\tdelay$, where \tdelay is the time required for the effects of \Rt to become apparent.
\tdelay is an important parameter of this analysis and will be discussed in paragraph~\ref{sec:td}. The models we take into consideration are:

\begin{itemize}
    \item Model-a $$ \Rtfit(t+\tdelay+1) = \Rt(t+\tdelay) \cdot \alpha_i $$ with $i$=yellow, orange, red. In this model, the effect of the NPIs  consists in a  multiplicative factor $\alpha$ on \Rt. This method has been adopted for instance in \cite{Flaxman, Brauner}.  Despite its popularity in literature, this model has severe limitations. The most important one is  that the values of \Rt can only asympotically tend to infinite (when $\alpha$ is greater than one) or to zero (when $\alpha$ is smaller than 1).
    \item Model-b \\  To overcome the limitations of Model-a, we introduce a model where \Rt can asymptotically converge to any finite value $R^*_{t,i}$:
    $$ \Rtfit(t+\tdelay+1) = \Rt(t+\tdelay) + \frac{\Rt(t+\tdelay)-R^*_{t,i}}{N} $$  The parameter $N$ defines the number of steps needed to reach \Rtstar.
    \item Model-c \\  The data of Fig.~\ref{fig:rt-ori} do not show discontinuities on the first derivative of \Rt, $\Rt^{'}$, in the regional trends. To guarantee the continuity of $\Rt^{'}$ we additionally refine Model-b in the following way:
    $$\Rtfit(t+\tdelay+1) =\Rt(t+\tdelay) + \frac{R^{'}_t(t+\tdelay) + \beta (\Rt(t+\tdelay)-R^*_{t,i})}{N} $$ 
    where the effects of the NPIs on the trend of \Rt are mediated by the  time derivative $\Rt^{'}(t+\tdelay)$; the weight $\beta$ is an additional degree of freedom in the fits.
\end{itemize}

We determine estimates of the parameters $\alpha_i$,  \Rtstar, \tdelay, N, $\beta$ by minimizing a $\chi^2$ defined globally over all considered regions, as described in the next section. 

% --------------------------------------------------------------------------------

\subsection{The time delay between NPI changes and their effect upon $\mathrm{R_{t}}$}
\label{sec:td}
The delay time \tdelay necessary to NPI changes to produce modifications in  \Rt trends, is determined by a scan on the overall $\chi^2$ computed for all the regions assuming Model-c in the period of time November 6$^{\mathrm{th}}$, 2020 -- January 15$^{\mathrm{th}}$, 2021.
The $\chi^2$ is defined as

%{\color{blue}(Perché non anticipiamo alla Sez. precedente questa definizione?
%Dovrebbe essere $R_t(i)$, no? Altrimenti vanno modificate le eq. precedenti.)}

\begin{equation}
    \chi^2 = \sum_{t=1}^{T}{\left(\frac{\Rt(t+\tdelay) - \Rtfit(t+\tdelay)}{\sigma_{\Rt(t+\tdelay)}}\right)^2}
\end{equation}
where  $t$  indicates the calendar days, $T$ is the total number of days in the considered period of time, $\sigma_{\Rt}$ is the error on \Rt. We have chosen Model-c because, as will be discussed in the following, it is the algorithm that best reproduces the data.
Moreover, we limited the period of time to January 15$^{\mathrm{th}}$, 2021 because, as also discussed in the following, it is the most stable period of time  before the introduction of the vaccination campaign and the spread of virus variants.

\begin{figure}[htb]
\centering
    \includegraphics[width=0.5\textwidth]{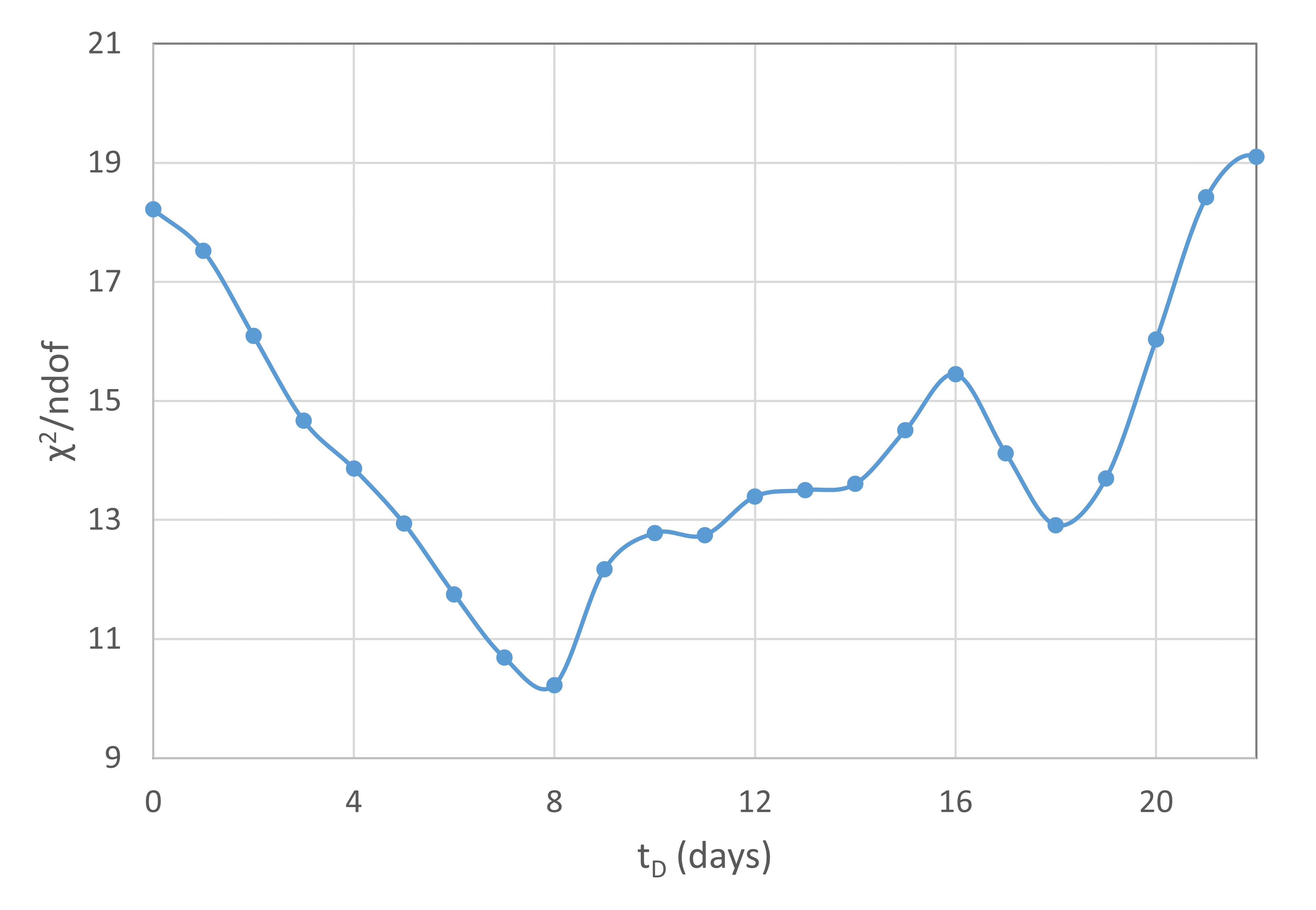}
    \caption{\footnotesize $\chi^2$ scan of the fit to \Rtstar with Model-c as a function of the delay time $\tdelay$. }
    \label{fig:chi2scan}
\end{figure}

The $\chi^2$ scan reported in Figure~\ref{fig:chi2scan}, obtained by varying the value of the time delay \tdelay in the range from 0 to 22 days,  indicates that an absolute minimum is reached for a delay of $\tdelay = 8$ days.
The $\chi^2$ distribution is quite broad and irregular; on the other hand, the
time delay is expected to have a quite broad distribution. Two contributions to the dispersion of \tdelay are the finite width of the generation time distribution, and the collection and processing time of molecular swabs that introduce a dispersion of 7.7 days as we estimated in~\cite{sintomatici}.

Since we assigned the Covidstat \Rt to the center of the time interval of the fit (8 days before the last day) and the Cori \Rt to the start of the time interval (7 days before the last day), the value of $\tdelay=8$ days implies that changes on NPI can be detected not earlier than 16 days (Covidstat) and 15 days (Cori) after their enforcement.

\begin{figure}[htb]
\centering
    \includegraphics[width=0.6\textwidth]{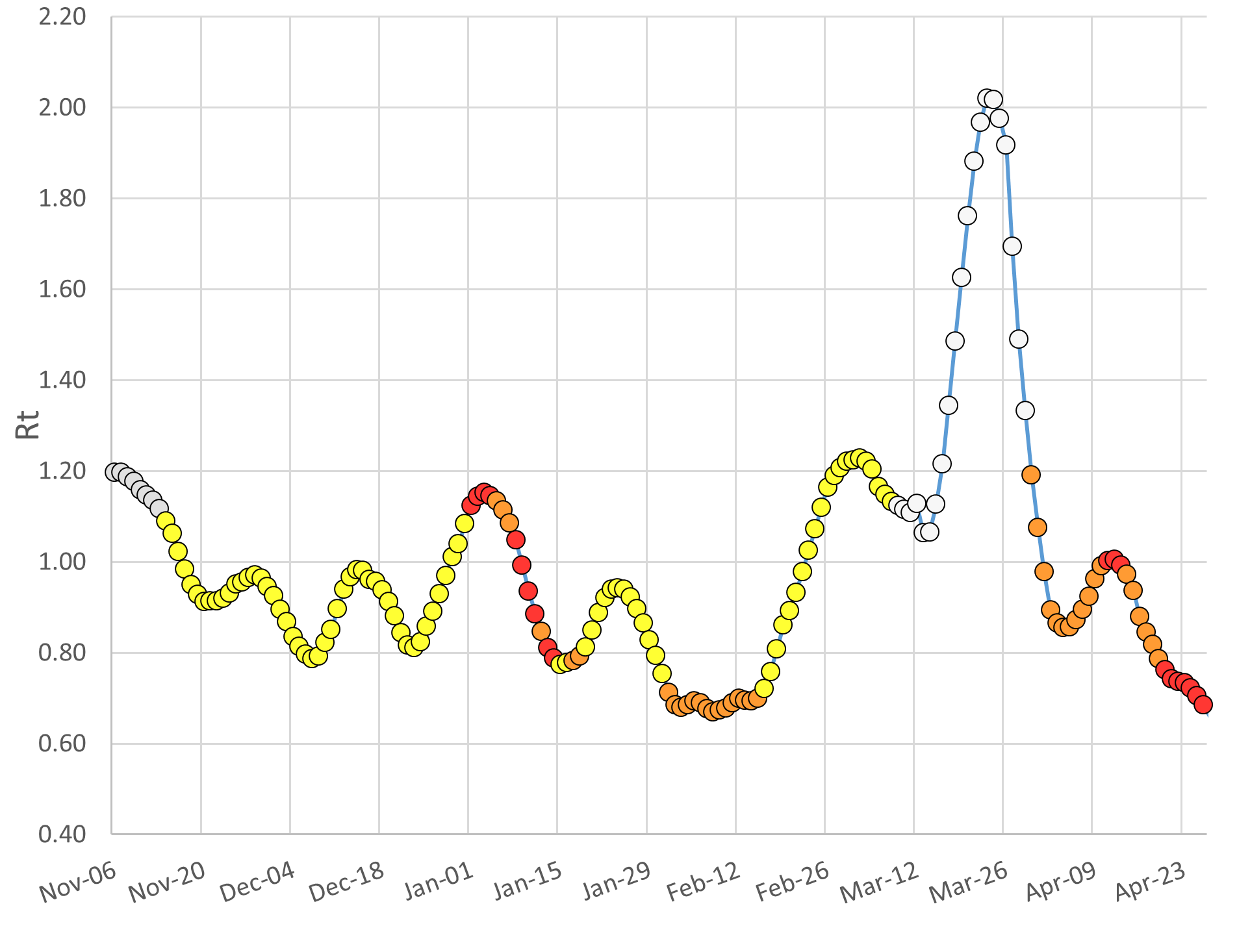}
    \caption{\footnotesize \Rt trend in Sardegna. The change to the white level was established on March 1$^{\mathrm{st}}$, 2021. Considering the required  delay of 8 days,     we display the change on March 9$^{\mathrm{th}}$, 2021. The steep ascent of \Rt begins on March 14$^{\mathrm{st}}$, 2021. The first 8 values of \Rt are colored in grey since they don't correspond to any of the three NPI levels here considered after having applied the 8 days time delay. %{\color{blue}Forse non ho capito, ma tu scrivi che il cambiamento a bianco avviene l'1 Marzo mentre in figura il bianco compare l'8 Marzo e il cambiamento di \Rt tra il 14 e il 15: temo di essermi perso... Cred che tu abbia spostato i colori di 8 giorni rispetto al vero inizio, ma questo il testo non lo dice, per cui ci si perde.}
    }
    \label{fig:Sardinia}
\end{figure}

As a cross-check, we display the trend of \Rt in Sardegna, which is not one of our selected regions, but it is the only one that achieved a white color assignment on March 1$^{\mathrm{st}}$, 2021, a decision which resulted in a pronounced rise of \Rt. The \Rt trend for Sardegna is shown in Figure~\ref{fig:Sardinia} and is in good agreement with our evaluation of $\tdelay=8$ days.

With an analogous $\chi^2$ scan, the number of steps $N$ for Model-b and Model-c is fixed to $N=9$.

% --------------------------------------------------------------------------------
\section{Effects of the three different NPI levels}

\begin{figure}[htb]
\centering
    \hspace*{0.4cm}\includegraphics[width=0.46\textwidth]{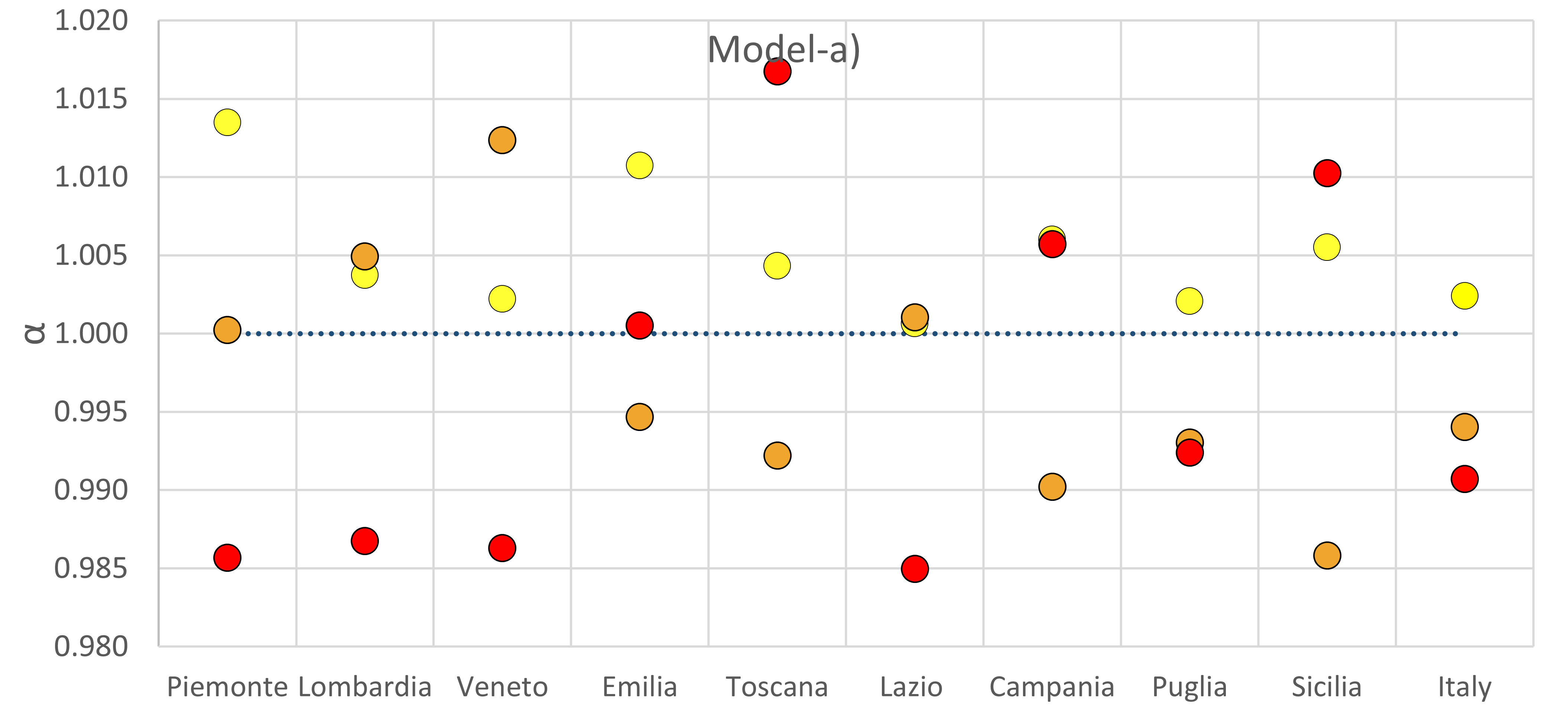}\hspace*{0.2cm}
    \includegraphics[width=0.46\textwidth]{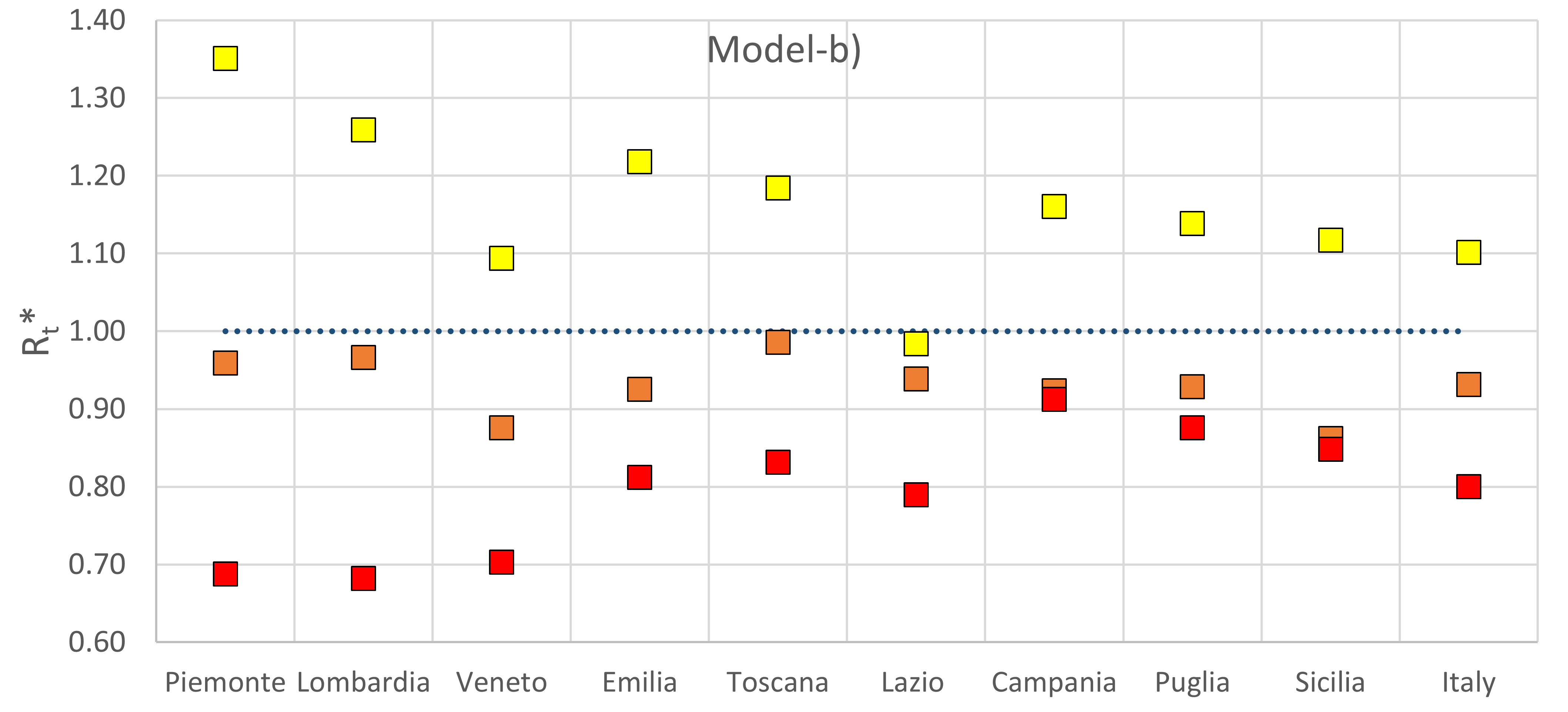}
    \vskip 0.4cm
    \includegraphics[width=0.79\textwidth]{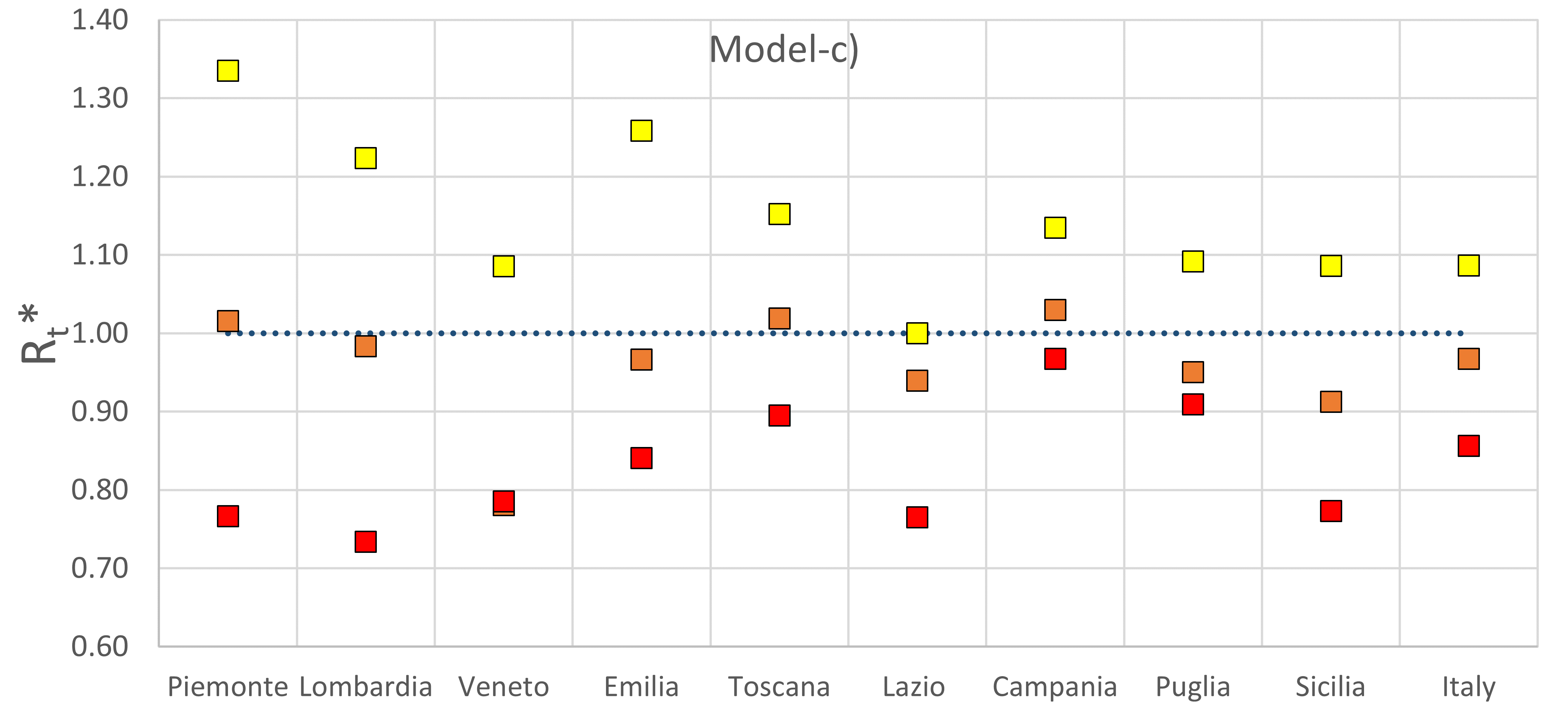}
    
    \caption{\footnotesize Results of the fits for the parameters of the Model-a (upper left plot; what is reported in dots  are the fitted estimates of the parameter $\alpha$, Model-b (upper right plot; the fitted estimates of \Rtstar are reported as squares) and Model-c (lower plot; the fitted estimates of \Rtstar are reported as squares) for all the regions and their combination (Italy columns). The colors of the fits, yellow, orange and red, correspond to the three NPI levels. Error bars are smaller than the markers and masked behind them. The horizontal blue dotted lines mark the threshold values $\alpha=1$ and $\Rtstar=1$. }
    \label{fig:results}
\end{figure}

We estimate the parameters of each model for 
all the individual italian regions and for their combination  (we will call it Italy from here on).
%assuming common parameter values for all Italy.
As a cross-check, we perform all the calculations by using both the Covidstat and Cori algorithms.
The results are displayed in Figure~\ref{fig:results}. The values for the Italian sample are reported in Table~\ref{tab:results}.

\begin{table}[htb]
\centering
\begin{tabular}{c|c|c}
NPI & \Rt Covidstat & \Rt Cori \\
\hline
\multicolumn{3}{c}{Model-a (parameter $\alpha_i$)}\\
\hline
Yellow      & $1.002 \pm 0.004$  &  $1.001 \pm 0.005$     \\
Orange      & $0.994 \pm 0.005$  &  $0.995 \pm 0.010$   \\ 
Red         & $0.991 \pm 0.005$  &  $0.987  \pm 0.018$  \\ 
\hline
\multicolumn{3}{c}{Model-b (parameter $R^*_{t,i}$)} \\
\hline
Yellow    & $1.102 \pm 0.0012$     &  $1.074 \pm 0.0010$ \\
Orange    & $0.931 \pm 0.0013$     & $0.928 \pm 0.0012$ \\ 
Red       & $0.800 \pm 0.0015$      & $0.762 \pm 0.0020$   \\ 
\hline
\multicolumn{3}{c}{Model-c (parameter $R^*_{t,i}$)}\\
\hline
Yellow     &  $1.086 \pm 0.0012$  & $1.078 \pm 0.0013$  \\
Orange     &  $0.967 \pm 0.0015$  &  $0.972 \pm 0.0014$ \\ 
Red         & $0.856 \pm 0.0016$    &  $0.804 \pm 0.0025$  \\ 
\end{tabular}
\caption{\footnotesize Fit values of the Model-a $\alpha_i$ parameters ($i$=yellow, orange, red) and of the $R^*_{t,i}$ parameters of Model-b and Model-c. }
\label{tab:results}
\end{table}
The overall $\chi^2/{\mathrm{ndof}}$ values are 33.4, 25.8, 20.0 for Model-a, Model-b, Model-c, respectively, as computed with the CovidStat \Rt algorithm for Italy and 36.2, 28.4, 24.3 for for the same models, using the Cori \Rt algorithm.
None of the $\chi^2$ is  fully satisfactory, but, as already discussed in Section~\ref{sec:Rtdata}, these models cannot claim to completely describe the course of the pandemic.

It's clear from the comparison of the $\chi^2$ values that Model-a does not provide a fit quality to the data comparable to Model-b and Model-c. The latter provides the best fit to the data and in the following we will consider it as our reference.

In all the italian regions the red set of NPI turns out to provide better containment than the orange set and this latter is better than the yellow set, well beyond the errors of the fits. Absolute numbers in the Italian regions can differ (as discussed in Section~\ref{sec:Rtdata}, different regions implemented additional and localised containment measures beyond the NPI considered here), but the average values in Italy are in good agreement with the regional trends.

The results indicate that the yellow NPI level  has been, in general, inadequate to guarantee the containment of the pandemic, since its fitted \Rtstar value is bigger than 1 for all cases (with the only exception of Lazio with an estimated $\Rtstar=1.0$). The orange NPI level produces trends dangerously close to $\Rtstar=1$, while the red NPI level guarantees a good containment of the pandemics. We recall that, as reported in Appendix, the main additional restrictions set by the red level were: further restrictions to mobility, closure of all the stores with few exceptions and suspension of all sport competitions.
Parameters computed with the CovidStat and Cori algorithms are in general in good agreement.

\begin{figure}[htb]
\centering
    \includegraphics[width=0.58\textwidth]{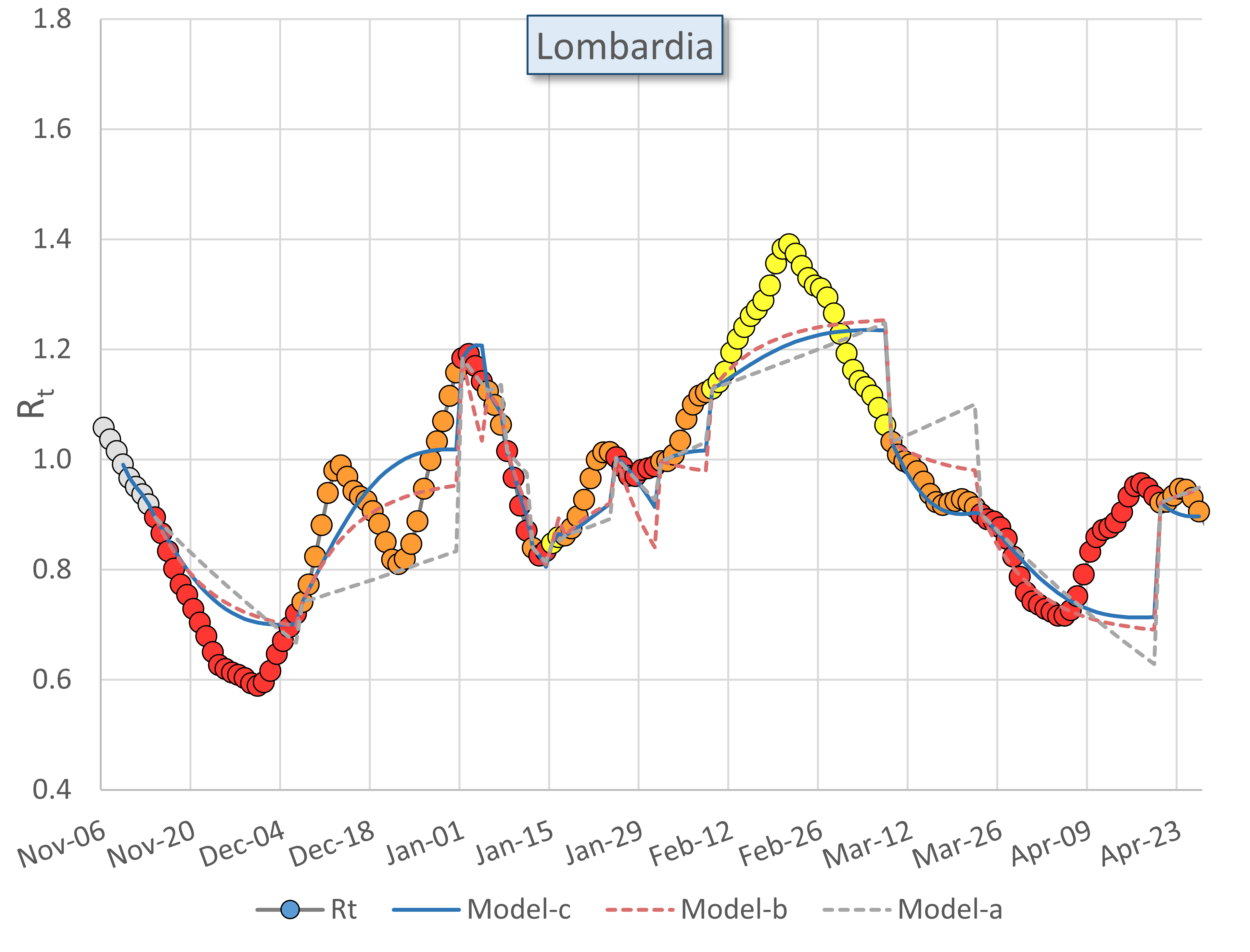}
    \caption{\footnotesize Estimated \Rtfit values in Lombardia computed with the Model-a (grey dotted curve), Model-b (pink dotted curve) and Model-c (blue curve) compared with the measured values of \Rt (colored dots). The colors of the \Rt dots correspond to the three NPI levels yellow, orange, red. They are delayed by 8 days as computed in the text. For this reason, the first 8 values of \Rt are colored in grey since they don't correspond to any of the three NPI levels here considered. We remind the fitted value $\Rtfit(t)$ is set equal to \Rt(t) at any change of NPI level}
    \label{fig:Lombardy-fits}
\end{figure}

\begin{figure}[htb]
\centering
    \includegraphics[width=0.32\textwidth]{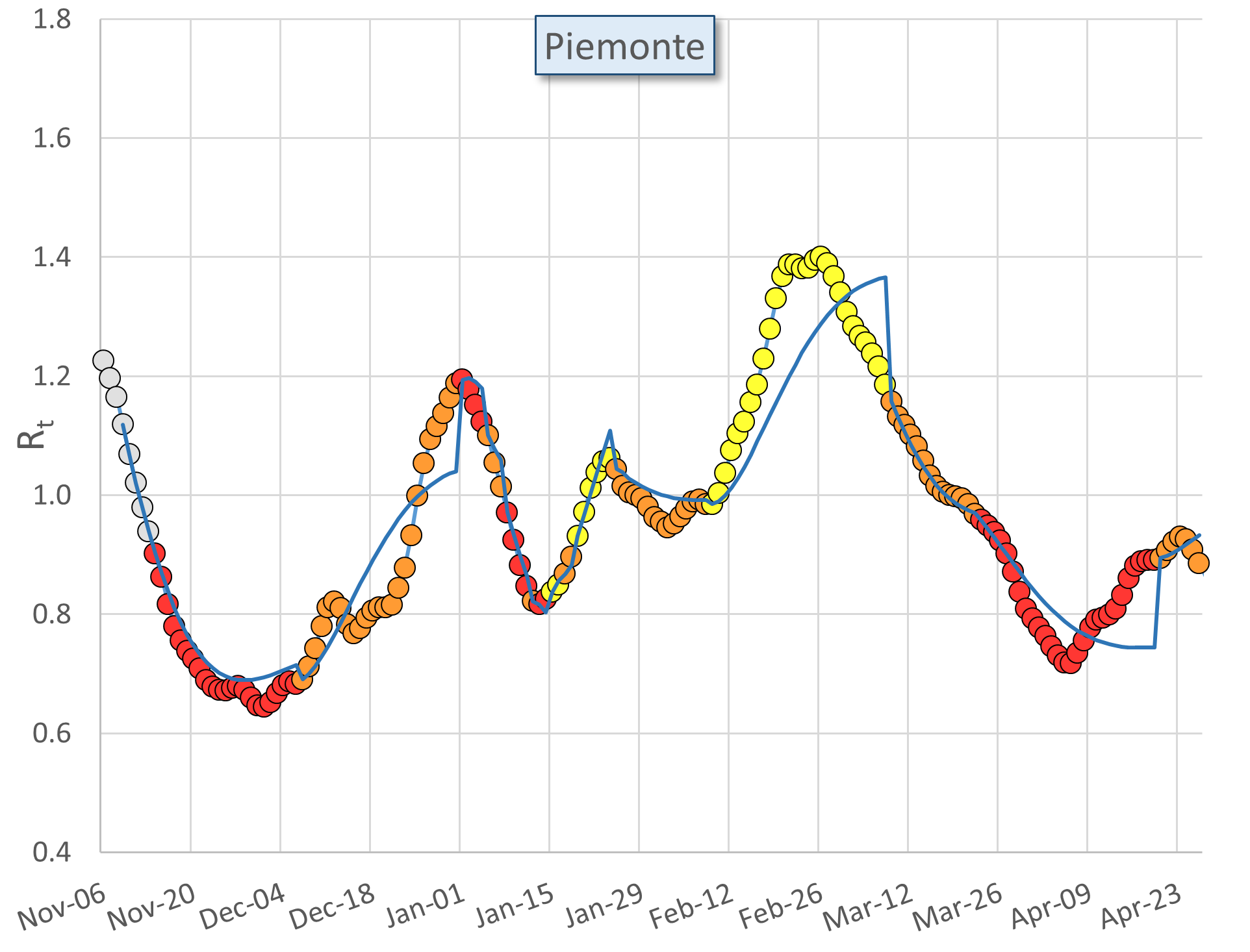}
    \includegraphics[width=0.32\textwidth]{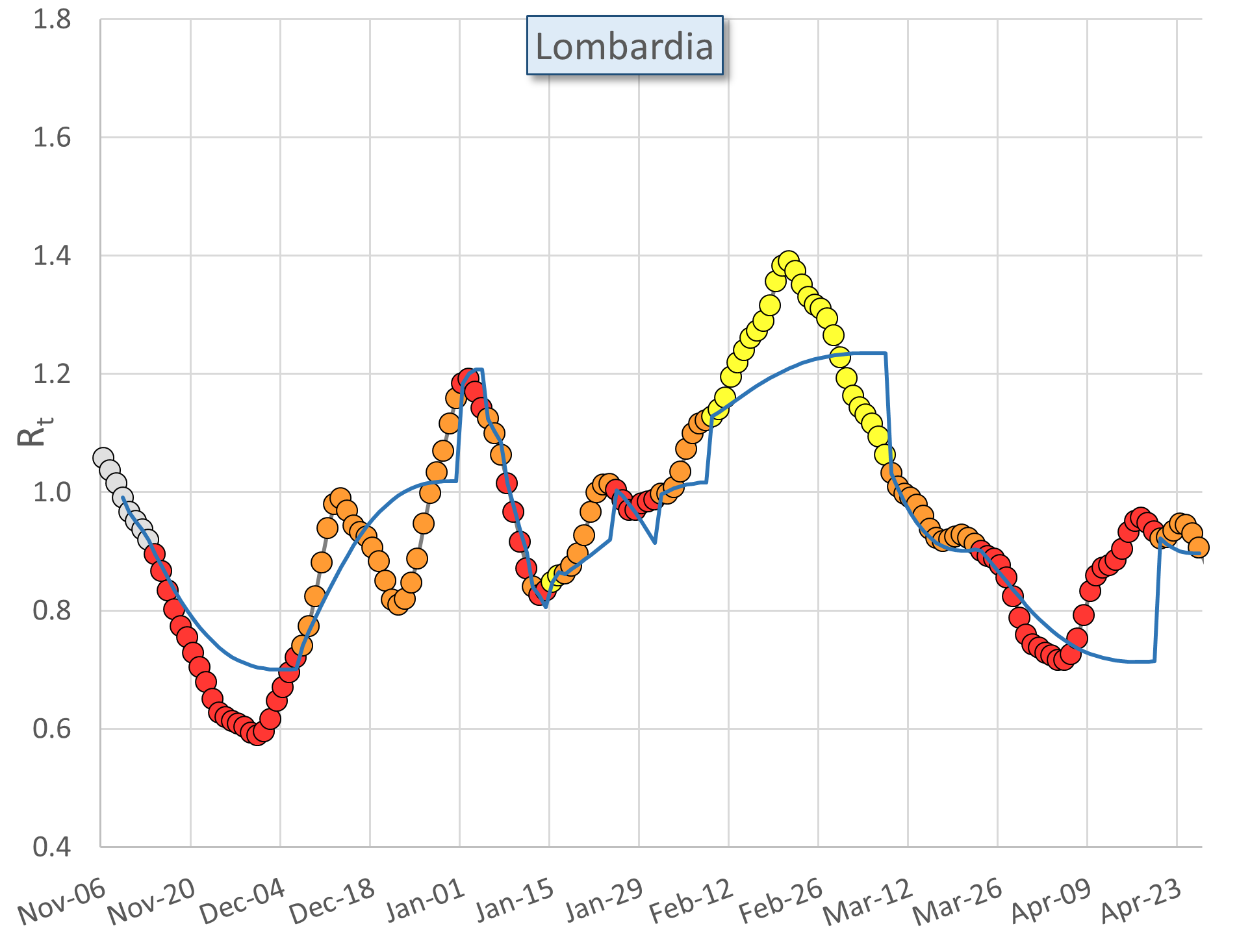}
    \includegraphics[width=0.32\textwidth]{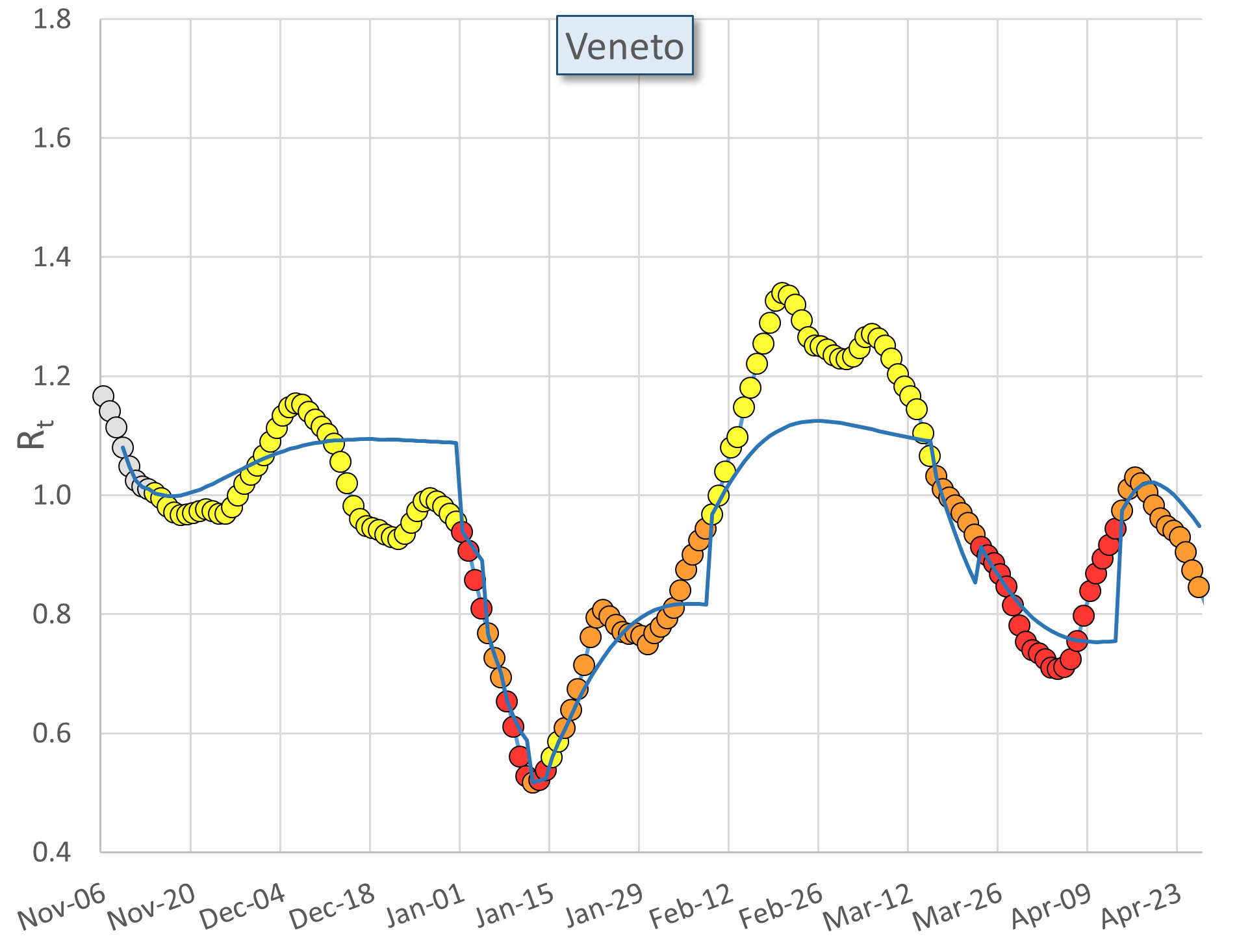}
    \includegraphics[width=0.32\textwidth]{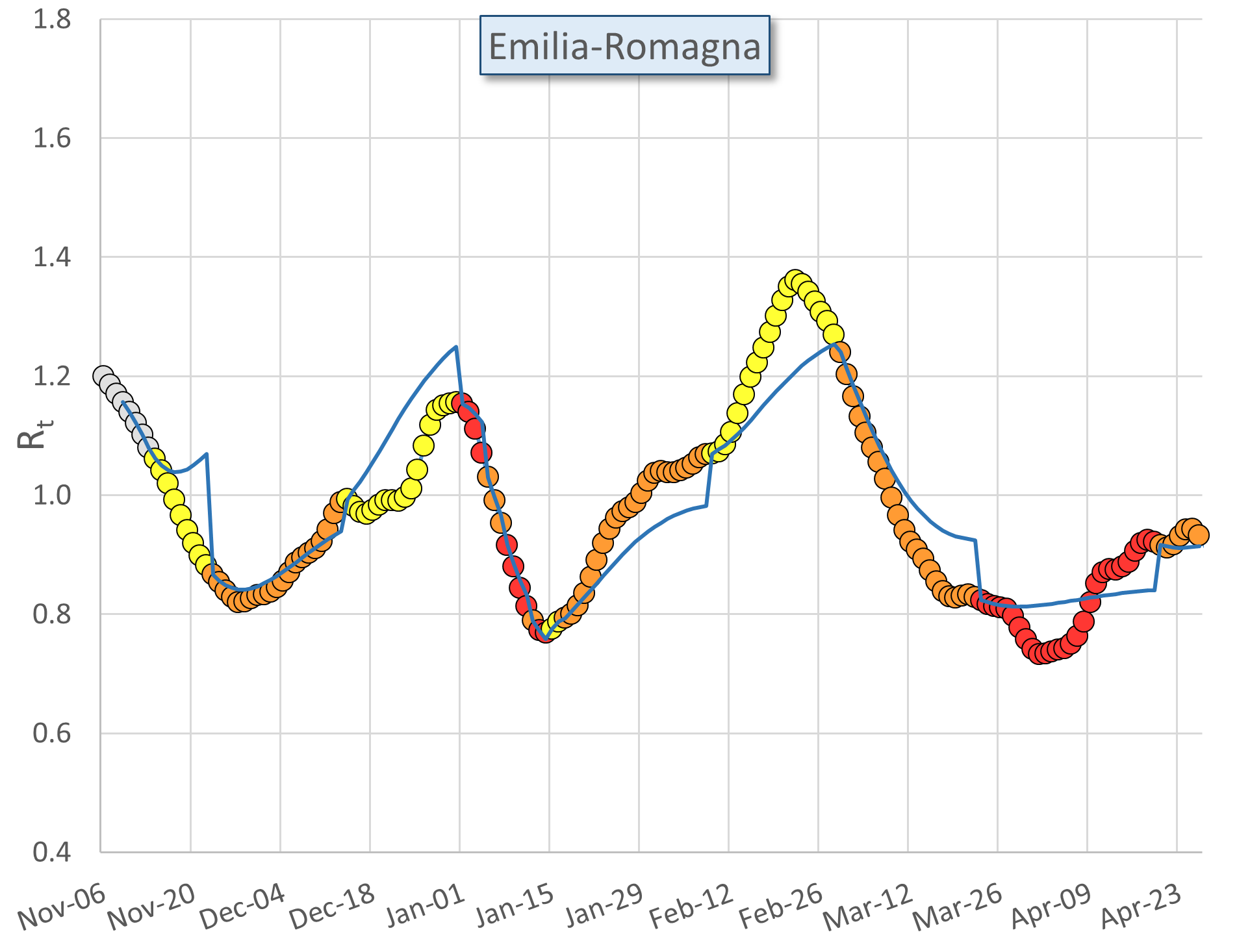}
    \includegraphics[width=0.32\textwidth]{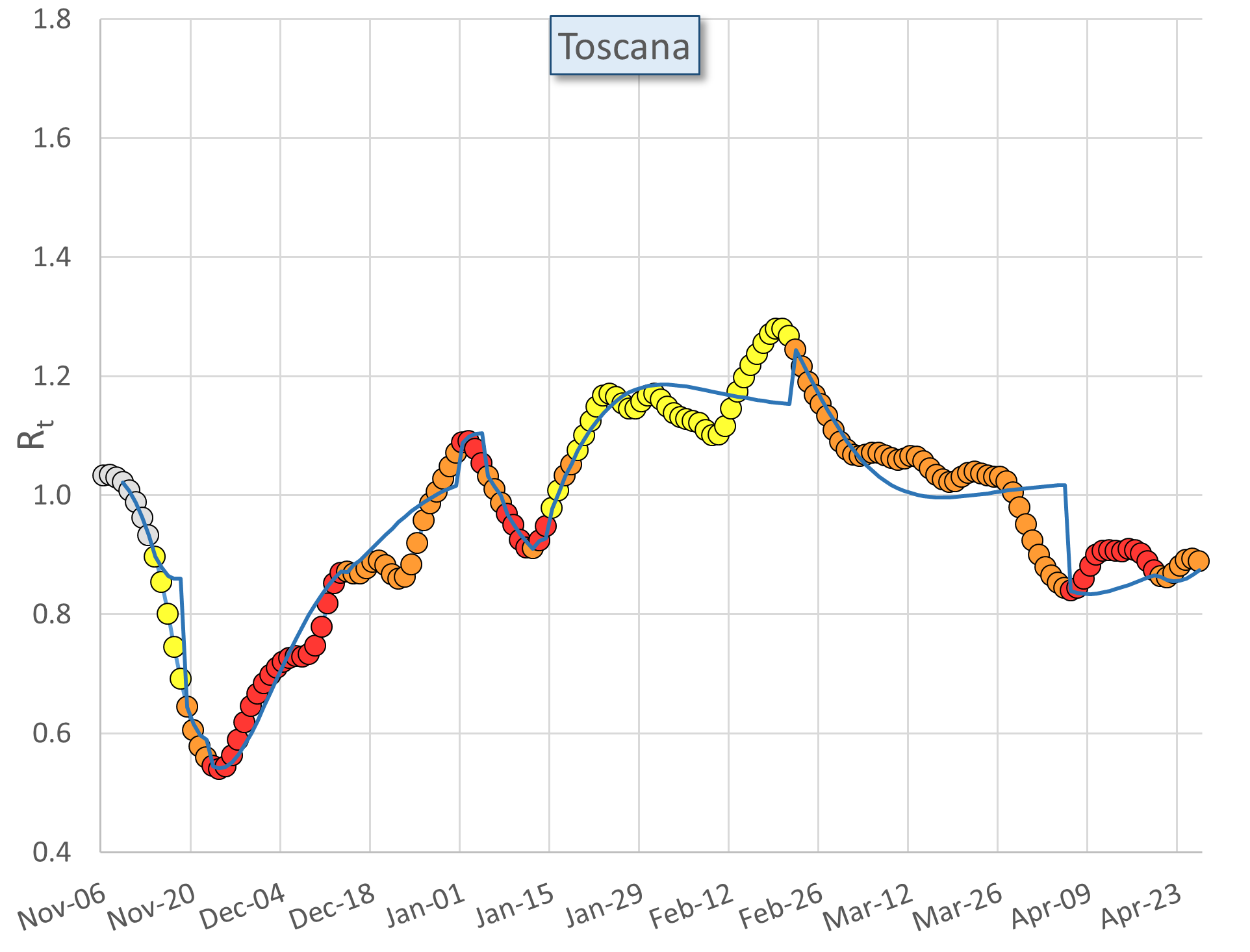}
    \includegraphics[width=0.32\textwidth]{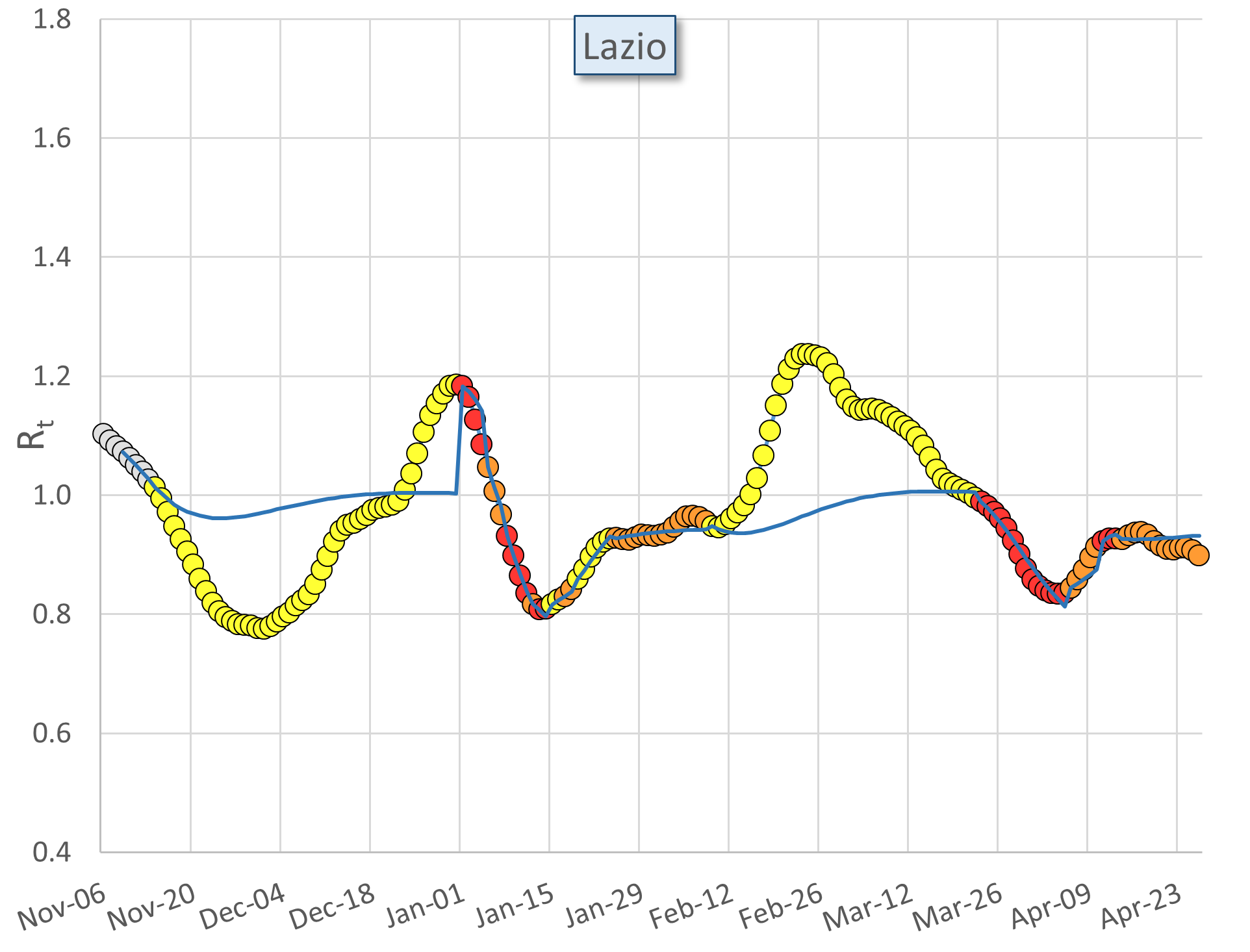}
    \includegraphics[width=0.32\textwidth]{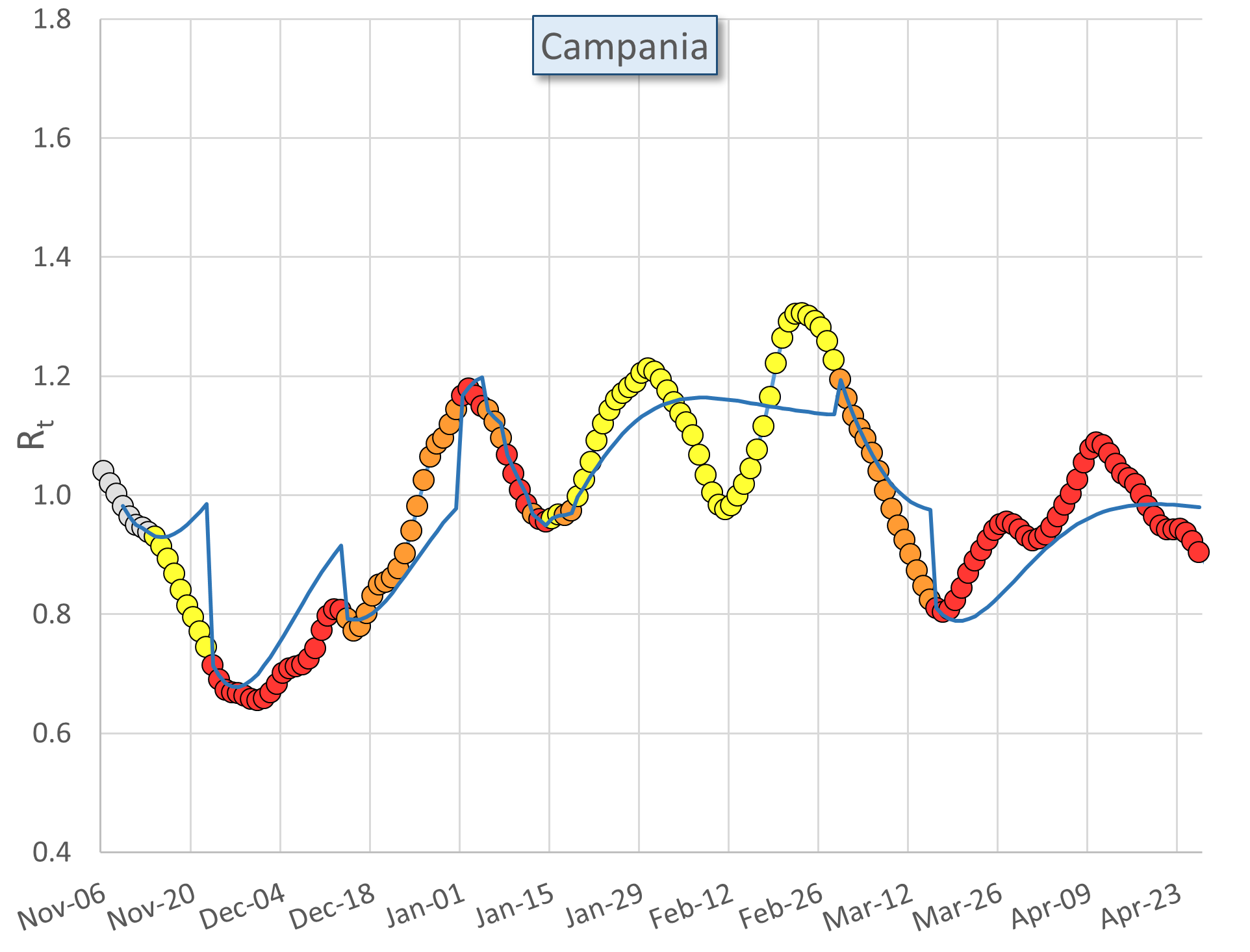}
    \includegraphics[width=0.32\textwidth]{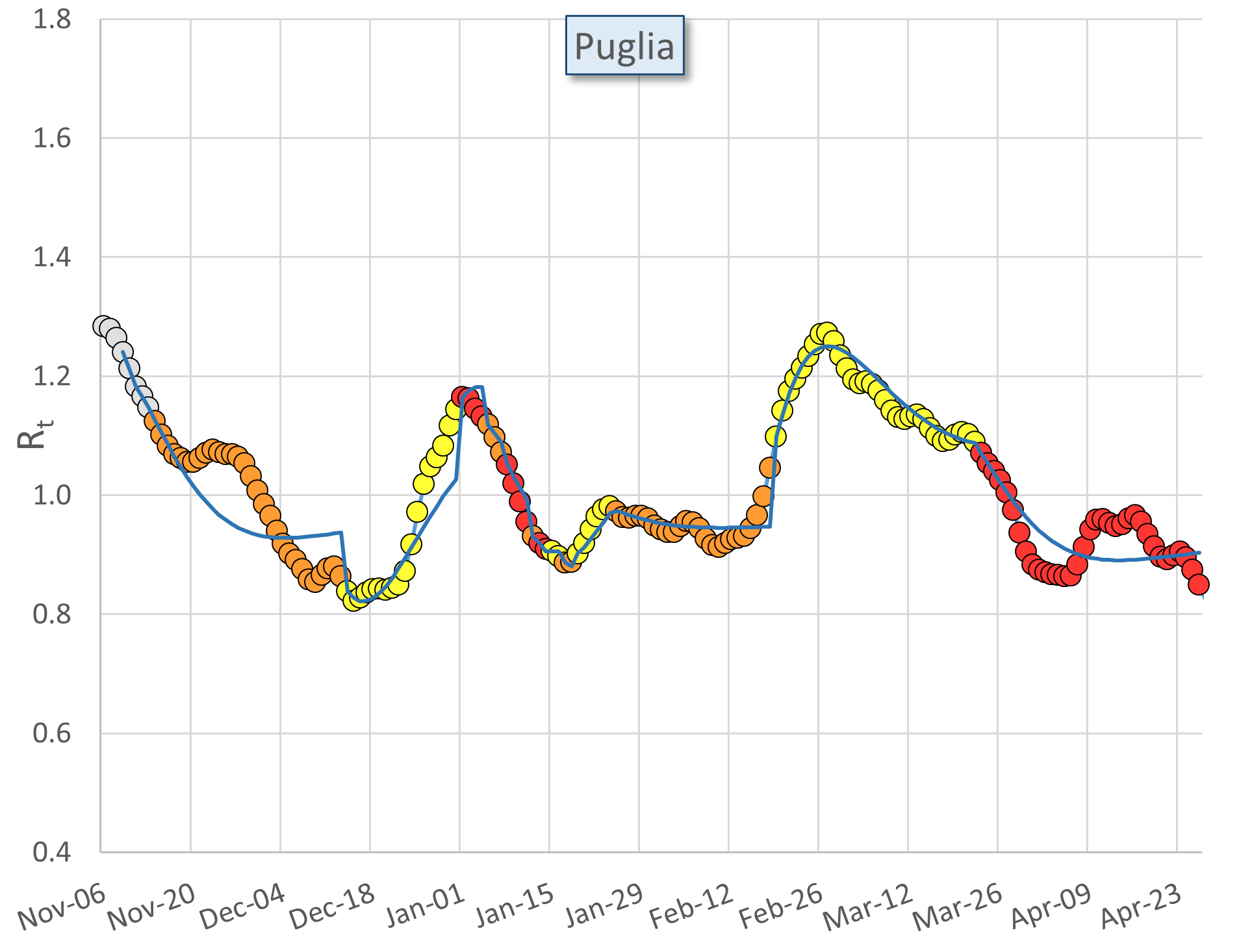}
    \includegraphics[width=0.32\textwidth]{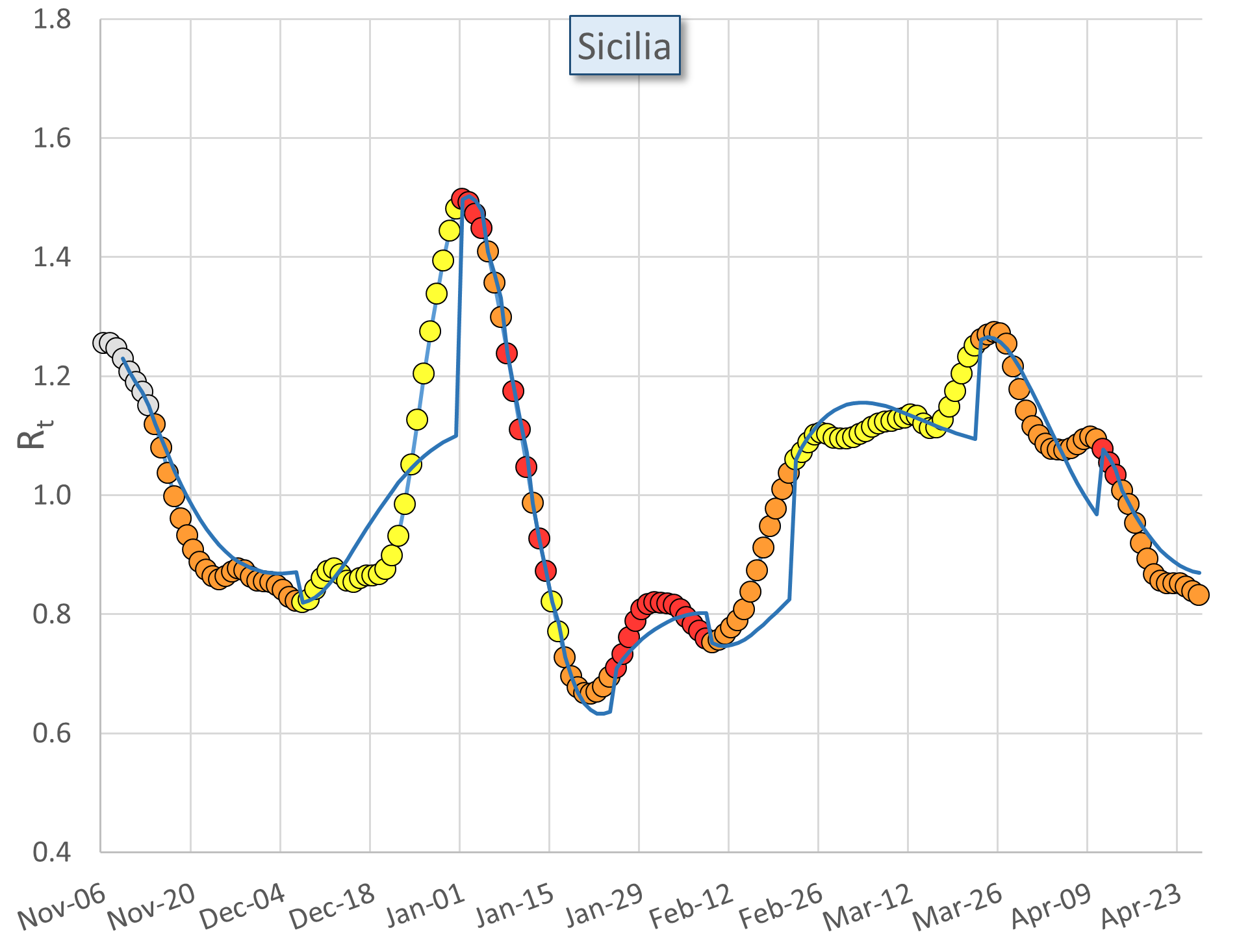}
    \caption{\footnotesize \Rt values as function of time in the nine considered regions. The colors of the dots correspond to the three NPI levels, yellow, orange, red after having applied a delay of $\tdelay=8$ days. For this reason, the first 8 values of \Rt are colored in grey since they don't correspond to any of the three NPI levels here considered.  The blue lines represent the output of the fits, \Rtfit, computed with Model-c for each individual region.}
    \label{fig:region-fits}
\end{figure}

We display in Figure~\ref{fig:Lombardy-fits} the fitted values  \Rtfit  computed with the three models superimposed to the  \Rt data in the case of Lombardia.  A visual inspection of Figure~\ref{fig:Lombardy-fits} confirms the outcome of the $\chi^2$ of the fits: Model-c is the model that better follows \Rt trends. It isn't fully satisfactory, but we have discussed in Section~\ref{sec:synch} that some features of the \Rt trends  are due to fluctuations of the data on positive swabs and we discussed in Section~\ref{sec:Rtdata} that the NPI we are modelling here are not the only factors that can influence the trends of \Rt.
In Figure~\ref{fig:region-fits} we display the fits to \Rt, \Rtfit computed with Model-c for all the regions. The colors of the \Rt dots in Figure~\ref{fig:region-fits} are different from those of Figure~\ref{fig:rt-ori} because here we have applied the delay time $t_D$. The \Rt trends differ significantly from region to region, it should be reminded that following the color assignments of Figure~\ref{fig:colori} the regions had different NPI assignments at different times. The resulting phenomenology results quite complicated. It's  significant that  Model-c on the average can follow the trends of \Rt in all the Italian regions.
% --------------------------------------------------------------------------------
\FloatBarrier
% --------------------------------------------------------------------------------
\section{Effects of the vaccination campaign and virus variants}
During the period of time considered in this study, two new important facts happened that could potentially impact on the trends of \Rt.
The vaccination campaign began in Italy early in January 2021. According to the data reported in the COVID-19 Opendata Vaccini repository~\cite{vaccini}, the fraction of the Italian population  vaccinated by January 15$^{\mathrm{th}}$, 2021 was 4.3\% (0\%) with the first (second) dose respectively, while, as April 26$^{\mathrm{th}}$, these fractions were increased to 20.1\% (8.3\%). 

On the other hand, virus variants (variants of concerns or VOC), specifically lineage B.1.1.7 (English variant), began to spread out in Italy in the same period: according to~\cite{ISS-varianti}, 86.7\% of the new cases in Italy were due to this variant at March 18$^{\mathrm{th}}$, 2021, while this percentage was 17.8\% at February 4$^{\mathrm{th}}$, 2021~\cite{ISS-varianti-1}. 

Both VOC and vaccinations  have a potentially important impact on the spread of the contagion,
%For these reasons, \st{that} {\color{blue}since these facts} should impact on \Rt in opposite ways, 
therefore we performed a fit with Model-c in two separate period of times: November 6$^{\mathrm{th}}$,2021 -- January 15$^{\mathrm{th}}$, 2021 and January 16$^{\mathrm{th}}$, 2021 -- April 26$^{\mathrm{th}}$, 2021 to measure a possible overall effect of  these two new factors in the progress of the pandemics.
The results of the fits are reported in Table~\ref{tab:varianti}.

\begin{table}[htb]
\centering
\renewcommand{\arraystretch}{1.1}
\begin{tabular}{c|ccc}
period of time & Yellow & Orange & Red \\
\hline
6 Nov, 2020 -- 15 Jan 2021 & $1.001 \pm 0.002$ & $0.968 \pm 0.003$ &  $0.802 \pm 0.004$ \\
16 Jan 2021 -- 26 Apr 2021 & $1.155 \pm 0.002$ & $0.974 \pm 0.002$ & $0.891 \pm 0.002$ \\
\end{tabular}
\caption{\footnotesize Fit values of the asymptotic values \Rtstar for the three NPIs levels yellow, orange and red, computed with Model-c (see text) in two different periods of time.}
\label{tab:varianti}
\end{table}

The combined effect of virus variants  and the vaccination campaign in the considered period of time,  resulted in a worsening of NPIs effectiveness. The yellow NPI level, for instance, worsened its effectiveness in containing the spread of contagion by 15\%. After this period of time, the effect of the  variant B.1.1.7 reached its maximum, while the vaccination campaign continues. So, hopefully, if no further variants will influence the  \Rt trends, the effect of the considered NPI  will improve.
% --------------------------------------------------------------------------------

% --------------------------------------------------------------------------------
\section{Conclusions}

The goal of this paper is the computation of the effectiveness of the three NPI called yellow, orange and red introduced in the Italian regions since November 6$^{\mathrm{th}}$, 2020.

We firstly synchronized the \Rt values with the dates in which the NPI changed in the Italian regions. The synchronization is a necessary step for any procedure that tries to compute changes in \Rt trends due to changes of NPI at given calendar dates. To this purpose we introduced two original methods, the first of which exploits the fact that our original Covidstat algorithm directly connects \Rt with the growth rate of an exponential fit.

We identified critical issues in the model often used in literature to quantify the effects of NPI measures. We introduced two methods to overcome them and the fit to the data shows that the two new models clearly outperform the original one. We selected the so called Model-c as a reference for all the results of the paper.

We demonstrated that our model is able to describe with just 4 parameters (3 asymptotic values \Rtstar and a weight parameter $\beta$)  the development of the pandemics in the 9 Italian regions considered in this paper. The development of the pandemics in the Italian regions is quite differentiated because they underwent different NPI restrictions at different times. The agreement between the model and the data is not complete because several local restrictive measures had been taken  and because the poor quality of the positive swabs data introduces important noise in the \Rt trends. We discussed these features in the paper.

The model provides a coherent picture where in all the Italian regions the red set of NPI results to have smaller \Rtstar values of the orange set and this latter has always \Rtstar smaller than the yellow set, well beyond the statistical errors. We find, that averaged over the nine most populated Italian regions, only the red NPI level can produce a significant reduction of \Rt, below the threshold value of 1. The orange level keeps \Rtstar around 1, while the yellow level  targets  \Rt values above one.

We measured a worsening of the containment effects in the period of time when the variant of concern lineage B.1.1.7 spread out in the country, even if, in the same period, the vaccination campaign started.

All the computations have been performed with two \Rt algorithms: the Covidstat algorithm that we developed~\cite{ourpaper} and the widely used standard Cori {\it et al.} algorithm~\cite{Cori}. We found a good agreement between the results obtained using the two algorithms.

\section{Acknowledgements}
We acknowledge the effort of ISS of making public the data of the symptomatic cases of COVID-19. The present work has been done in the context of the INFN CovidStat project that produces an analysis of the public Italian COVID-19 data. The results of the analysis are published and updated daily on the website 
{\tt https://covid19.infn.it/}. The project has been supported in various ways by a number of people from different INFN Units. In particular, we wish to thank, in alphabetic order: Stefano Antonelli (CNAF), Fabio Bredo (Padova Unit), Luca Carbone (Milano-Bicocca Unit), Francesca Cuicchio (Communication Office), Mauro Dinardo (Milano-Bicocca Unit), Paolo Dini (Milano-Bicocca Unit), Rosario Esposito (Naples Unit), Stefano Longo (CNAF), and Stefano Zani (CNAF). We also wish to thank Prof. Domenico Ursino (Universit\`a Politecnica delle Marche) for his supportive contribution.

% --------------------------------------------------------------------------------
\newpage
\appendix
\setcounter{secnumdepth}{0}
\section{Appendix}
\label{app:risks}

A brief description of the rules in place for the four NPI levels denominated as White, Yellow, Orange, Red is reported in table~\ref{tab:app}.

Following the Ministerial Decrees of October 20$^{\mathrm{th}}$, 2020 \cite{schools-1} and March 13$^{\mathrm{th}}$, 2021 \cite{schools-2}, nursery and elementary schools had been always open, except the period March, 3$^{\mathrm{rd}}$, 2021 - April 2$^{\mathrm{nd}}$, 2021 when they closed. High schools and Universities were suggested to organize distance teaching, with some degree of freedom at local level. On April 6$^{\mathrm{th}}$, 2021 it has been decided to bring the distance teaching at 50\% and on April 26$^{\mathrm{th}}$, 2021 this percentage become 70\% to 100\%.  middle schools (three years, 12 to 14 years old students) had different implementations in the different regions, the most common was to follow for the first year the rules of elementary schools and for the following two years the rules of high-schools.
\vskip 0.5cm
\begin{table}[htb]
    \renewcommand{\arraystretch}{1.1}
{
    \centering
    \footnotesize
 \begin{tabular}{|p{0.8\textwidth}|}
  \hline
   \vskip 0.2cm\centerline{\normalsize\textbf{White}}
   \begin{itemize}
        \item Requirement to wear masks outdoors
        \item Social distancing of 1 meter
        \item Curfew: decision taken at the regional level, Sardegna chose a curfew from 11:30PM to 5:00AM. 
    \end{itemize}\\
 \hline 
    \vskip 0.2cm\centerline{\normalsize\textbf{Yellow}} 
    All the restrictions of White, plus:
    \begin{itemize}
       \item 
        Curfew from 10:00PM to 5:00AM.
        \item
        Shopping centers closed on holidays.
        \item
        Museums and exhibitions closed.
        \item
       Distance education for high schools; in-person teaching for preschools, elementary schools, and middle schools. Universities closed, except for laboratories 
       \item
       Public transportation must travel with max 50\% occupancy
       \item
       Coffee shops and restaurants open until 6:00PM. Take-away allowed until 10:00PM. Home deliveries without restrictions
       \item
       Suspension of activities of amusement arcades, betting shops.
       \item
       Pools, gyms, theaters, and movie theaters remain closed. Open sports centers.
       \end{itemize}\\
    \hline
    \vskip 0.2cm\centerline{\normalsize\textbf{Orange}}
    All the restrictions of Yellow, plus:
    \begin{itemize}
        \item 
        Prohibited movement in and out of regions and municipalities (with some exceptions)
        \item
        Coffee shops and  restaurants closed 7 days a week.Take-away allowed until 10:00PM. Home deliveries without restrictions
    \end{itemize}\\
     \hline
    \vskip 0.2cm\centerline{\normalsize\textbf{Red}}
    All the restrictions of Orange, plus:
    \begin{itemize}
        \item 
    It is forbidden any movement, even within its own municipality, except for reasons of work, needs and health.
     \item
        Closed all stores, except supermarkets, grocery stores, newsstands, tobacconists, pharmacies, laundries, hairdressers.
    \item
    Suspended all sports competitions except those recognized as being of national interest. Closed sports centers.
    \end{itemize}\\
 \hline
 \end{tabular}
 }
     \caption{\footnotesize Brief description of the rules in place for the four NPI levels denominated as White, Yellow, Orange, Red.}
    \label{tab:app}
 \end{table}
 
% --------------------------------------------------------------------------------
\end{document}